\documentclass{aa}  

\usepackage{graphicx}
\usepackage{caption}
\usepackage{txfonts}
\usepackage{ulem}
\usepackage{color}
\usepackage{array} 
\usepackage{multirow} 
\usepackage{makecell} 
\usepackage{hyperref}
\usepackage{natbib}
\begin{document}

   \title{Earth-like planets hosting systems: Architecture and properties}


   \author{Jeanne Davoult
          \inst{1}
          \and
          Yann Alibert\inst{2}
          \and
          Lokesh Mishra \inst{3}
          }

   \institute{Space research \& Planetary Sciences (WP), Universität Bern,
              Gesellschaftsstrasse 6, 3012 Bern, Switzerland\\
              \email{jeanne.davoult@unibe.ch}
            \and
             Centre for Space and Habitability (CSH), Universität Bern, Gesellschaftstrasse 6, 3012 Bern, Switzerland\\
             \and
             IBM Research, R\"uschlikon, Z\"urich, Switzerland
             }

   \date{Received 24 January 2024; accepted 24 July 2024}

 
  \abstract
   {The discovery of Earth-like planets is a major focus of current planetology research and faces a significant technological challenge. Indeed, when it comes to detecting planets as small and cold as the Earth, the cost of observation time is massive. Understanding in what type of systems Earth-like planets (ELPs) form and how to identify them is crucial for preparing future missions such as PLATO, LIFE, or others.
   Theoretical models suggest that ELPs predominantly form within a certain type of system architecture. Therefore, the presence or absence of ELPs could be inferred from the arrangement of other planets within the same system.}
   {This study aims to identify the profile of a typical system that harbours an ELP by investigating the architecture of systems and the properties of their innermost detectable planets. Here, we introduce a novel method for determining the architecture of planetary systems and categorising them into four distinct classes. We then conduct a statistical study to identify the most favourable arrangements for the presence of an ELP.}
{Using three populations of synthetic planetary systems generated using the Bern model around three different types of stars, we studied the `theoretical' architecture (the architecture of a complete planetary system) and the `biased' architecture (the architecture of a system in which only detectable planets are taken into account after applying an observation bias) of the synthetic systems. To describe a typical system hosting an ELP, we initially examined the distribution of ELPs across different categories of architectures, highlighting the strong link between planetary system architecture and the presence of an ELP. A more detailed analysis was then conducted, linking the biased architecture of a system with the physical properties of its innermost observable planet to establish the most favourable conditions for the presence or absence of an ELP in a system.}
   {First, using synthetic systems, we successfully reproduce the distribution of observed multi-planet systems within the five different architectural classes. This demonstrates the relevance, at the system level, of populations of the synthetic systems derived from the Bern model and the observational bias applied. Secondly, the biased architectures (with observation bias) correspond for the most part to the theoretical architectures (without bias) of the same system. Finally, the biased architecture of a system, studied in conjunction with the mass, radius, and period of the innermost detectable planet, appears to correlate with the presence or absence of an ELP in the same system.}
  {We conclude that the detections of ELPs can be predicted thanks to the already known properties of their systems, and we present a list of the properties of the systems most likely to host such a planet.}


   \maketitle
%
\section{Introduction}
In the last decade, the number of known planetary systems with multiple planets has increased to such an extent that the study of the architecture of these systems has leapt forward. While the study of exoplanets was led by the study of these bodies as isolated objects, the increase in the amount of observed multiplanet systems has allowed the emergence of a new facet of exoplanetary science: the study of planetary systems as a whole, the correlations between planets \citep{Millholland2017,Weiss2018, Weiss2022}, and the comparisons not between planets, but between systems (e.g. \cite{Alibert2019, Gilbert2020}); for a recent review of the study of planetary systems, see \cite{Zhu2021}.
There are even different approaches to treating these planetary systems as a unified subject of study: the dynamical stability of these systems (e.g. \cite{LaskarPetit2017, Yeh2020, Stalport2022}, the link between the composition of the systems and the stellar and protoplanetary disc properties (e.g. \cite{Petigura2018,Mulders2021,Adibekyan2021}), as well as the architectural study of planetary systems (\cite{Lissauer2011,Millholland2017,Weiss2018,Mishra2023} among others). Considering these planets as interconnected is to emphasise that they were formed by the same mechanisms at the same time and that their evolution has kept the marks of the evolution of the others. Hot Jupiters are a perfect example: small rocky planets are rarely found in systems with internal giant planets, a likely trace of the chaotic past of this type of system \citep{Latham2011,Steffen2012}. Indeed, in many cases, the type-II migration of a giant planet will lead to a dynamical reorganisation of the system by the ejection of small rocky planets from the inner system. In contrast, systems consisting of Earths or super-Earths forming a chain, a formation also called `Peas-in-a-Pod' \citep{Weiss2018}, is widespread and most likely the consequence of the synchronous formation and migration of these planets in their protoplanetary discs, trapping them in resonances, followed by an instability phase \citep{Izidoro2017,Goldberg2022, Batygin2023}. Other studies have shown that these peas-in-a-pod configurations are also energy-optimising states for low-mass systems \citep{Adams2019,Adams2020}. These chains of terrestrial planets are generally clustered within a few fractions of au (Trappist-1 \citep{Gillon2017} is a perfect example), and the presence of planets close to the habitable zone of a star often reveals the presence of a planet in this habitable zone. The architecture of systems, consisting of the arrangement of planets in a system, is then considered to be a remnant of the formation process of systems.
These links between formations and architectures directly motivate the present study. Connecting architecture to a type of formation means that planets in the same system should have correlated properties. These correlations could allow us to obtain information on planets we have yet to detect.\\
The architecture of planetary systems has already been studied in \cite{Mishra2023, Mishra2023b}, whose work has defined a framework that classifies any system in an architecture type based on the mass of the planets in that system. In this framework, four types of architecture can describe any system with at least two planets: \textit{Similar}, \textit{Anti-Ordered}, \textit{Ordered}, and \textit{Mixed}. The connection between Earth-like planets (ELP in the rest of the paper) and the \textit{Similar} class has been established.\\
With upcoming missions such as PLATO \citep{Rauer2014, Rauer2016} or LIFE \citep{LIFE1}, whose objective is to discover ELPs within the habitable zone of their stars, it is crucial to prepare the detections. 
Studies have estimated that the rate of occurrence of ELPs in the habitable zone of main-sequence stars ranges from less than 0.27 \citep{Hsu2019, Bergsten2022} to between 0.58$^{\text{+0.73}}_{\text{-0.33}}$ and 0.88$^{+\text{1.28}}_{-\text{0.51}}$ planets per star for the optimistic habitable zone \citep{Bryson2021}. This large uncertainty is mainly due to the very small number of ELPs presently detected. However, only four habitable zone ELPs are estimated within 10~pc around G or K stars with 95\% confidence \citep{Bryson2021}. Understanding the types of systems in which these planets form and predicting which systems have the highest probability of hosting an ELP will ultimately save observation time and ensure successful detections. 
This work proposes a new method, largely inspired from \cite{Mishra2023}, of establishing four architectural classes and subsequently using this classification to expose the correlations between the architecture of a system, the properties of the innermost detectable planet --- which also give us information on the conditions of formation of the planetary system ---  and the presence of an ELP in a system. With the aim of using such correlations to distinguish between planetary systems with and without ELPs among systems where no ELPs have been detected yet, a theoretical observational bias is applied to synthetic populations of planetary systems. Observational bias allows us to study the theoretical and biased architectures of the same synthetic system, which correspond to the architectures of the system with all the modelled planets (theoretical) or with only the planets that can be observed, while accounting for the limitations and biases inherent in the observational methods (biased).\\
Section \ref{sec:synthpop} describes the global model of planetary formation, and the synthetic planetary system populations used in this study, and the observational bias applied to study the observable properties of these systems. Section \ref{sec:features} presents the properties of the systems taken into account in the correlation study and discusses their relevance. Section \ref{sec:correlations} presents the results of the study of correlations between system properties and the presence of an ELP. Finally, we discuss and conclude in section \ref{sec:conclu}.

\section{Synthetic populations of planetary systems} \label{sec:synthpop}
   To investigate the correlations between the properties of planets within the same system, particularly with the presence of an ELP, this study utilises synthetic planetary system populations generated by the Bern model. These synthetic planetary systems give us access to a large dataset of complete systems in which we know all the planets, particularly whether there is an ELP. To study the observable properties of these systems, we apply an observation bias to them. The following subsections describe the model, the populations of planetary systems used in this study, and the detection bias.
   \subsection{The Bern model and synthetic populations of planetary systems} \label{subsec:bernmodel}
\begin{figure*}[t]
    \centering
   	\includegraphics[width=6cm]{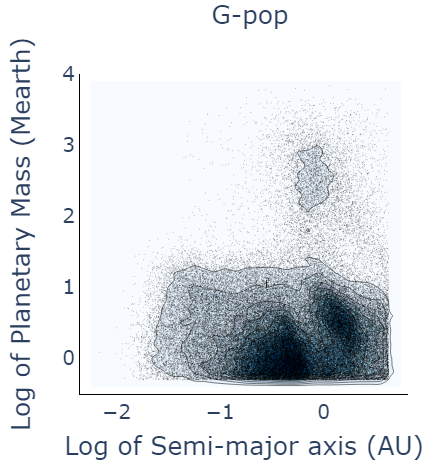}
   	\includegraphics[width=6cm]{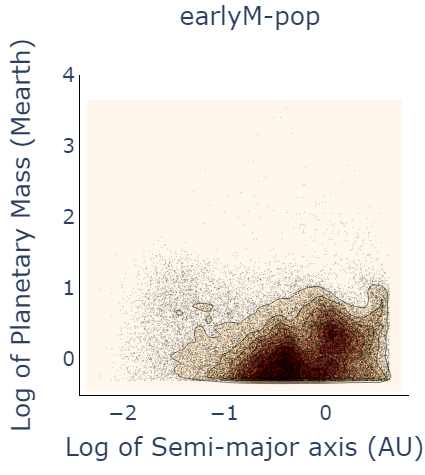}
   	\includegraphics[width=6cm]{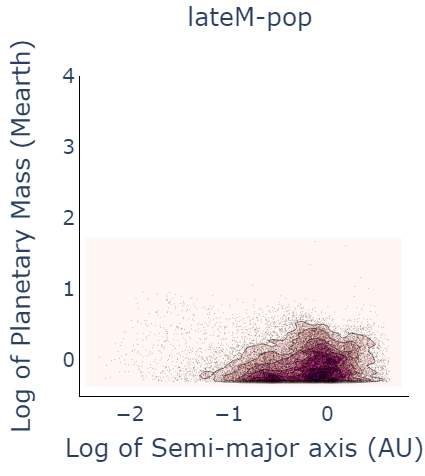}
   	
   	\caption{Representation of G-, earlyM-, and lateM-pop. The X-axes represent the log of the semi-major~axis in AU-unit, and the Y-axes represent the log of masses in M$_{\oplus}$-unit. The colour bars correspond to the concentration of planets.}
   	\label{fig:pop}
    \end{figure*}

   The Bern model is a global model of planetary system formation and evolution using the population synthesis method. The model is briefly reviewed here; readers interested in more details can refer to the recent NGPPS series of papers in which the model is described in great detail in \cite{NGPPS1} and analysed in \cite{NGPPS2,NGPPS3, NGPPS4, NGPPS5,NGPPS6}. The historical development of the Bern model is reviewed in \cite{Benz2014, Mordasini2018}.\\
   The Bern model is based on the planetesimal-based core-accretion paradigm \citep{Pollack1996}. This model simulates the formation and evolution of planets from lunar-mass protoplanetary embryos embedded in protoplanetary discs around stars of various masses. During the formation phase of the model (20 Myr), the simulated planets form via solid and gas accretion, and migration and N-body interactions are included. After 20 Myr, the gas disc is depleted, and only the thermodynamical evolution of the synthetic planets is computed for another 10 Gyr.\\
   When computing population synthesis, some of the initial conditions are fixed, such as the stellar mass, the gas disc viscosity ($\alpha$ = 2 $\times$10$^{-3}$), the planetesimals' size (radius = 300 m) and density (rocky 3.2 g cm$^{-3}$, icy 1g cm$^{-3}$), and the initial density profile of the gas and planetesimals discs via power laws \citep{Veras2004}:
   \begin{equation}
   	\label{gas}
   	\Sigma_g(t=0)=\Sigma_{g,0}\left(\frac{r}{r_0}\right)^{-\beta_g}exp\left(-\left(\frac{r}{R_{char}}\right)^{2-\beta_g}\right)\left(1-\sqrt{\frac{R_{in}}{r}}\right),
   \end{equation}
	where r$_0$ = 5.2 au is the reference distance, $\beta_g$ = 0.9 the power-law index \citep{Andrews2010}, R$_{char}$ the characteristic radius, and R$_{in}$ the inner edge of the disc. 
   The rest of the initial conditions are drawn at random following distributions constrained by observations or theoretical arguments and are different for each system:
   \begin{itemize}
   	\item the initial mass of the gas disc, M$_g$ \citep{Beckwith+1996}
   	\item the external photo-evaporation rate, $\dot{M}_{wind}$ \citep{Haisch2001}
   	\item the dust-to-gas ratio, f$_{D/G}$=M$_s$/M$_g$ (M$_s$, is the mass of the solid disc) \citep{Murray2001,Santos2003}
   	\item the inner edge of the gas disc, R$_{in}$ \citep{Mordasini2009}
   	\item the initial location of the embryos \citep{Mordasini2009}.
   \end{itemize} 
   By changing the mass of the central stars, the modelisation of planetary formation changes, and so do the populations. The modelisation of the protoplanetary disc directly depends on the central star's mass. The initial aspect of the disc is shaped by M$_{disc}$, R$_{in}$, and R$_{char}$, all three of which vary by changing the mass of the central star. These changes are discussed in detail in \cite{NGPPS4} but weshall give a brief analysis of the changes encountered in three populations.\\
   As the planetary formation is linked with the central star's mass, populations around three different stars are expected to vary. The frequency of planet types is not the same in the three populations, and it is therefore crucial to consider different star types in this study.
   The synthetic populations of planetary systems used in this work are three populations of systems computed with 20 embryos initially embedded in the protoplanetary disc, and they are represented in a semi-major~axis-mass diagram in Fig. \ref{fig:pop}:
   \begin{itemize}
   	\item G-pop: 14505 systems around solar mass stars
   	\item earlyM-pop: 9681 systems around 0.5 solar mass stars
   	\item lateM-pop: 9963 systems around 0.2 solar mass stars
   \end{itemize}
   For each population, the smallest (M$_p$ < 0.5~M$_{\oplus}$) or farthest (period > 15 years) objects were removed to ensure that only completely formed objects and comparable bodies were kept. Indeed, we assume that planets with a period greater than 15 years are not likely to be confirmed as exoplanets with most of our current detection methods, and the exclusion of planets with masses smaller than 0.5~M$_{\oplus}$ is discussed in Appendix \ref{App:exclusion}.\\ 
    A clear distinction between the three populations is the reduction in the number of giant planets, or more generally massive planets, with decreasing stellar mass (predicted by \cite{Laughlin2004}, see also tables \ref{tab:frac_syst} and \ref{tab:multicity_special} for a quantitative analysis).\\ 
    A more detailed analysis can be made on populations for the following different categories of planets:
    \begin{itemize}
        \item ELPs: 0.5 < M$_p$ < 3~M$_{\oplus}$ in the temperate zone around their stars
        \item super-Earths: 3 < M$_p$ < 10~M$_{\oplus}$
        \item Neptunian planets: 10 < M$_p$ < 30~M$_{\oplus}$
        \item sub-giant planets: 30 < M$_p$ < 100~M$_{\oplus}$
        \item giant planets:  M$_p$ > 100~M$_{\oplus}$
    \end{itemize}
   The category ELP in this paper refers to planets of about the mass of the Earth (0.5 $\le$ M$_p$ $\le$ 3~M$_{\oplus}$) evolving in the `temperate zone' of the star. This temperate zone is extended compared to the habitable zone around a star, which enables us to work with a larger quantity of data. It corresponds to a zone extending between 160 and 510~K, with the equilibrium temperature calculated as follows:
   \begin{equation}
   	T_{eq}[K] = 279 \times sma[AU]^{-1/2}L_{\star}[L_{\odot}]^{1/4}.
   \end{equation} 
    On average, the temperate zones within the G-pop span a range of 0.39 to 3.90 AU from the star. In the earlyM-pop, this range narrows to 0.25 to 2.52 AU; in the lateM-pop, it further decreases to 0.15 to 1.48 AU. This defined range of ELP in the temperate zone of its star will serve as the category of planets of interest studied throughout this work.\\
    Tables \ref{tab:frac_syst} and \ref{tab:multicity_special} provide a comprehensive overview of the quantitative analysis of the five different types of planets within the three populations. Table \ref{tab:frac_syst} specifically presents the distribution of each planet type within each population. Notably, most systems in the G-pop host an ELP (60\%), which increases even further to 74\% in the earlyM-pop. In contrast, only 40\% of systems in the lateM-pop possess an ELP. Despite having the lowest abundance of ELPs, the lateM-pop is important in the search for ELPs. These planets are comparatively easier to detect around low-mass stars, which are the most prevalent in our solar neighbourhood.\\
    The prevalence of other planet types decreases as the central star's mass decreases, and very few systems possess giants when the central star's mass diminishes to 0.5 or 0.2~M$_{\odot}$. This significant difference can be attributed to the decreasing mass of the protoplanetary disc as it correlates with the central star's mass, thereby impacting the availability of materials required for forming such planets. It is, therefore, crucial to account for different types of stars.\\
    
   \begin{table}[h]
   	\caption[]{Fraction of systems with specific planetary types for the three different populations}
   	\begin{tabular}{lccc}
   		\hline
   		& G-pop & earlyM-pop & lateM-pop \\
   		\hline
   		ELP   & 0.6 & 0.74 & 0.4\\
   		super-Earth & 0.57 & 0.44 & 0.05\\
   		Neptunian & 0.3 & 0.14 & < 0.01\\
   		sub-giant & 0.11 & 0.01 & < 0.01\\
   		giant & 0.19 & < 0.01 & 0 \\
   		\hline
   		\label{tab:frac_syst}
   	\end{tabular}
    \end{table}
    Table \ref{tab:multicity_special} provides insights into the multiplicity of each planet type within systems that harbour that particular type of planet, referred to as the `mean multiplicity'. This metric helps determine whether a planet type tends to occur in groups. The mean multiplicity is calculated as the number of type-X planets in the entire population divided by the number of systems with at least one type-X planet ($n_{sys\cup X}$). This metric cannot be lower than unity.
    \begin{equation}
	\bar{n_X}=n_X/n_{sys\cup X}
    \end{equation}
    The planets in ELP and super-Earth categories tend to occur in groups, consistent with observations of Earth or super-Earth clusters \citep{Millholland2017, Millholland2021, Weiss2018, Weiss2022}. On the other hand, more massive planets such as Neptunian or sub-Jovian planets are more likely to appear as solitary instances within their respective types, with mean multiplicities ranging from 1 to 1.42 and 1 to 1.18, respectively. Interestingly, giant planets, absent in our lateM-pop, are scarce and solitary (less than 1\% of systems with a giant and mean multiplicity of 1.11) within the earlyM-pop. However, within the G-pop, they can appear in groups of two (mean multiplicity of 1.67).\\
    These results indicate distinct behaviours of systems based on the type of planets they contain. It is common to find rocky planets (Earths or super-Earths) in groups, while giant planets tend to be more isolated within their systems, which confirms the need to account for architecture.\\
    \begin{table}[h]
    \caption[]{Mean multiplicity of each planetary type in the three populations. For example, on average, there are 3.13 ELPs for each G star with at least one ELP.}
    \begin{tabular}{lccc}
    \hline
	& G-pop & earlyM-pop & lateM-pop \\
	\hline
    ELP   & 3.13 & 3.28 & 2.60\\
    super-Earth & 2.43 & 2.09 & 1.20\\
    Netunian & 1.42 & 1.24 & 1\\
    sub-giant & 1.18 & 1.22 & 1\\
    giant & 1.67 & 1.11 &  N.A.\\
    \hline
    \label{tab:multicity_special}
    \end{tabular}
    \end{table}

    \subsection{Observation bias} \label{subsec:bias}
\begin{figure*}[t]
   	\centering			
   	\includegraphics[width=5cm]{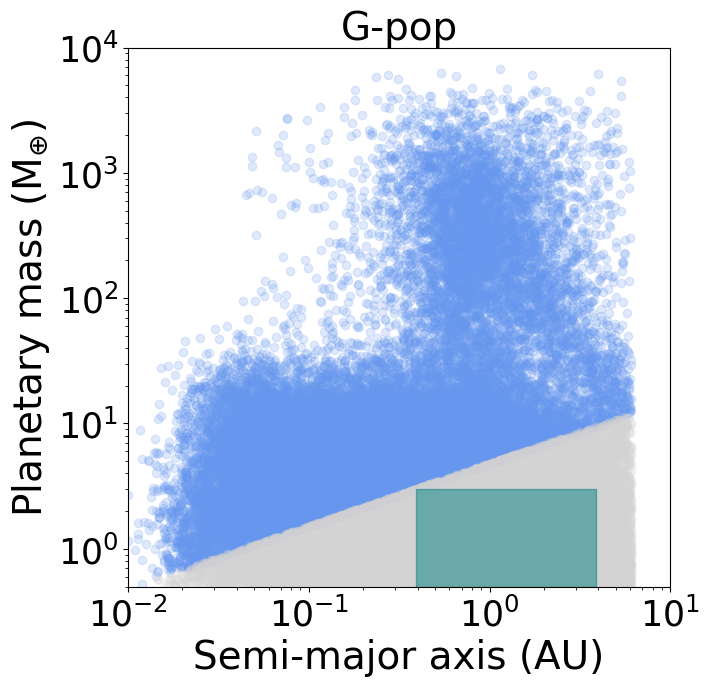}
   	\includegraphics[width=5cm]{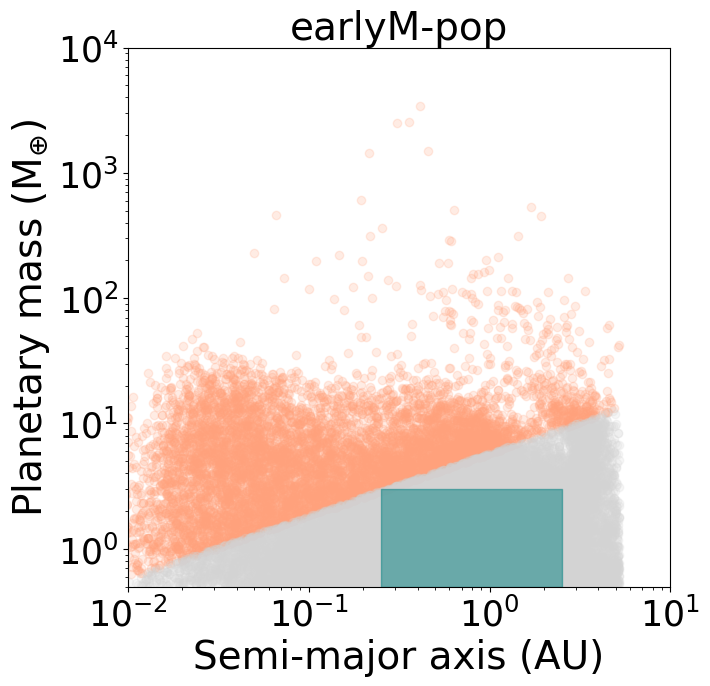}
   	\includegraphics[width=5cm]{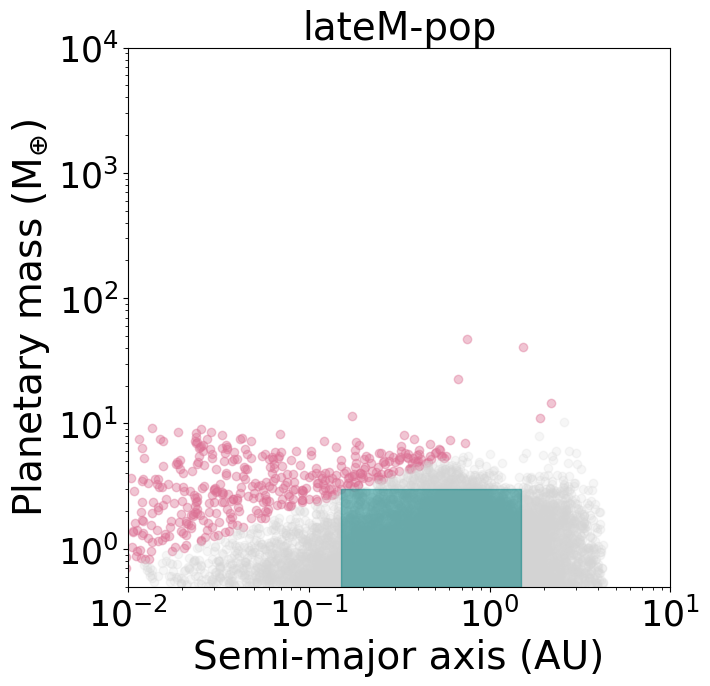}
   	
   	\caption{Representation of biased G-, earlyM-, and lateM-pop . X-axes represent the log of the semi-major~axis in AU-unit, and the Y-axes represent the log of masses in M$_{\oplus}$-unit. The coloured dots represent visible planets with radial velocity semi-amplitudes exceeding the threshold, while the grey dots represent invisible planets with radial velocity semi-amplitudes below the threshold. The green boxes represent the definition of an ELP adopted in this study. }
   	\label{fig:pop+ELP}
   \end{figure*}

   In order to explore the correlations between the presence of an ELP and observable features in a planetary system, one must apply an observational bias onto the synthetic planetary systems.
   The biases of radial velocity and transit methods , the two major methods for detecting exoplanets, limit our ability to observe all planets in a system and only allow us to detect the closest and largest ones. Consequently, these observational biases may affect the visible architecture of the synthetic populations.\\
   We employ a synthetic observational bias applied to the population mentioned above to investigate this case. An RV-related observational bias was chosen since the architecture classes are defined in terms of planetary mass (see section \ref{subsec:archi}). This bias involves defining a threshold in RV semi-amplitude, below which a planet is considered undetectable, and the remaining planets above the threshold form the biased system. This process drastically alters the system's architecture compared to the theoretical system's architecture (i.e. the one without bias).\\
   The radial velocity semi-amplitude of a star induced by a planet is calculated as follows:
   \begin{equation}
   	K_{RV}[m.s^{-1}] = 0.6395 \cdot P[days]^{1/3}\cdot M_p[M_{\oplus}]\cdot M_{\star}[M_{\odot}]^{-2/3}
   \end{equation}
   with
   \begin{equation}
   	P[days] = 365.25\cdot sma[AU]^{3/2}\cdot M_{\star}[M_{\odot}]^{-1/2}
   \end{equation}
   The detection threshold used in the study varies for each population and is defined by the limit of detection of an ELP. Indeed, we are examining two specific cases: systems without ELPs and systems where ELPs remain undetected. We are excluding cases where ELPs are detected. This approach assumes our observational bias prevents us from detecting ELPs, thereby setting the limit of our observational capability. Setting aside the scenario of systems with a detected ELP is not a strong assumption, as there are currently only 24 ELP-hosting systems (0.5\%) among the more than 4300 observed systems (from exoplanet.eu catalogue, June 2024), following our definition of an ELP (see Sect. \ref{subsec:bernmodel}).\\
   To exclude all ELPs from our systems, the detection threshold for a population is calculated based on the most massive ELP located closest to the star (top-left corner of the green boxes in Fig. \ref{fig:pop+ELP}). Since the range in periods of the habitable zone (green boxes in Fig. \ref{fig:pop+ELP}) varies with the mass of the star, and the radial velocity semi-amplitude also varies as M$^{-2/3}$, the detection threshold increases with decreasing stellar mass, resulting in the thresholds displayed in Table \ref{tab:threshold retained} for the three populations. The different values of the bias for each population (see Table \ref{tab:threshold retained}) are higher or equivalent to the authentic bias. For G-pop, the bias is relatively low. However, it remains within achievable ranges of values (with, e.g. ESPRESSO \citep{Pepe2021}). Alternatives to this bias are discussed in the discussion section (see Sect. \ref{sec:conclu}).\\
   
   \begin{table}[h]
   	\caption[]{RV semi-amplitude thresholds retained for each population.}
   	\begin{tabular}{lr}
   		\hline
   		\textbf{Population} & \textbf{threshold retained} \\
   		G-pop   & 0.43 m.s$^{-1}$\\
   		earlyM-pop & 0.76 m.s$^{-1}$\\
   		lateM-pop & 1.55 m.s$^{-1}$\\
   		\hline
   		\label{tab:threshold retained}
   	\end{tabular}
   \end{table}
   For an overview of the three retained thresholds, Fig. \ref{fig:pop+ELP} depicts the three populations, each in a diagram of semi-major axis and of planetary mass. It highlights the detectable planets forming the new systems (coloured dots), the undetectable planets (grey dots), and the definition of an ELP adopted in this study (represented by the green boxes).
   It is worth noting that the observational bias in RV real observations is more complex than the one used in this study. The bias varies from one instrument to another, depending on the stellar activity, observation segmentation, or even human bias. The observational bias used here is too simplistic to be fully accurate, but it has the advantage of being particularly simple to reproduce.\\

   These three populations of synthetic planetary systems, whose theoretical and biased forms are known, serve as the foundation for this study. They allow for the study of correlations between undetectable planets, particularly ELPs, and detectable planets, enabling predictions for future observations. In the rest of the paper, theoretical refers to the complete synthetic system without observational biases (with all planets with M$_p$~>~0.5~M$_{\oplus}$), and biased refers to the incomplete systems, which only include detectable planets when considering observational biases.
\section{Features considered for their correlation with the presence of an Earth-like planet}\label{sec:features}
    \subsection{Planetary systems' architectures} \label{subsec:archi}
    The planetary system architecture is essential in searching for ELPs. \cite{Zhu&Wu2018} found a correlation between an Earth's presence and a cold giant in the same system. In contrast, an anti-correlation is found between the presence of a hot Jupiter and the presence of inner terrestrial planets (e.g. \cite{Steffen2012}). In their study of system architecture classes, \cite{Mishra2023} highlight that their ‘Similar’ class harbours the most ELPs in the habitable zone of their stars. Thus, the planets' arrangement in their system is a key point for predicting the presence of an Earth in a system.
    
    \subsubsection{An architecture framework inspired by \cite{Mishra2023}} \label{subsubsec:framework}
    \cite{Mishra2023,Mishra2023b} propose a new way to classify planetary system architecture through four classes and their associated formation pathways. They define their architectural classes using two coefficients, one of which is the coefficient of similarity C$_s$:
    \begin{equation*}
        C_s=\frac{1}{n-1}\sum^{i=n-1}_{i=1}\left( log \frac{m_{i+1}}{m_i}\right),
    \end{equation*}
    with $m_i$ the mass of the i$^{th}$ planet and $n$ the total number of planets in the system. Due to a cancellation of the logs, this coefficient captures the variation of the mass between the first and the last planet. In this work, we are interested in the variation across the entire system. Therefore, we have opted to employ a new method for categorising system architectures using Principal Component Analysis (PCA) that leads to four architecture classes largely inspired by the ones defined in \cite{Mishra2023}. We designed this new framework to include a \textit{Low-mass} class instead of \textit{Similar}. Even though the \textit{Similar} class contains almost only systems with low-mass planets, theoretically, it also includes systems with massive planets of similar masses, which we want to avoid. The \textit{Anti-ordered}, \textit{Ordered}, and \textit{Mixed} classes keep the same names as those of \cite{Mishra2023} because they were largely inspired by that work and express the same notion of mass arrangement.\\
    In order to define our new framework, we first visually classified 200 systems with at least two planets. This initial `human' classification provided a framework we mathematically described afterwards. We defined the \textit{Low-mass} architecture to correspond to systems in which all planets are of relatively low mass and the maximum value for `low mass' has been set at 20~M$_{\oplus}$, which approximately corresponds to the critical core mass for runaway accretion as found in \cite{Bodenheimer1986} (M$_{crit}$ $\in$ [10~M$_{\oplus}$,30~M$_{\oplus}$]). See Appendix \ref{App:bornes} for more details.\\
    The other architectures relate to systems with at least one planet more massive (one planet with M$_p$ > 20~M$_{\oplus}$), and their classification is based on the arrangement of these planets' masses. To differentiate them, we applied a PCA to planetary systems using two properties: the x-axis corresponds to the logarithmic distance from the star, and the y-axis corresponds to the logarithmic masses of the planets. The PCA provides the first two components (axes) with the highest dispersion (variance). These two components represent each system by an ellipse, where the major axis is the first component of the PCA, and the minor axis is the second component. These two axes can then characterise the architecture of the system: the direction in which the ellipse points (the slope of the first component S(C$_1$)) determines whether masses increase or decrease with distance from the star, and the width of the ellipse (the variance of the second component V(C$_2$)) describes the amount of correlation between planetary properties or lack thereof in a system. Consequently, we defined the \textit{Anti-Ordered}, \textit{Ordered}, and \textit{Mixed} architectures based on the PCA results. The \textit{Ordered} (\textit{Anti-Ordered}) architecture corresponds to a system with a positive (negative) slope of the first component and low variance of the second component (V(C$_2$)< 0.2). In contrast, the \textit{Mixed} architecture corresponds to a system with high variance of the second component (V(C$_2$) > 0.2), regardless of the slope of the first component. In summary, our classes are defined as follows, with max(M$_p$) representing the mass of the heaviest planet, S(C$_1$) denoting the slope of the first component, and V(C$_2$) indicating the variance of the second component:

   	\begin{itemize}
   		\item \textbf{Low-mass}:  $max$(M$_p$)~$\le$~20~M$_{\oplus}$ \\
   		\item \textbf{Anti-ordered}:  $max$(M$_p$)~>~20~M$_{\oplus}$ and V(C$_2$)~<~0.2 and S(C$_1$)~$\le$~0 \\
   		\item \textbf{Ordered}:  $max$(M$_p$)~>~20~M$_{\oplus}$ and V(C$_2$)~<~0.2 and S(C$_1$)~>~0 \\
   		\item \textbf{Mixed}:  $max$(M$_p$)~>~20~M$_{\oplus}$ and V(C$_2$)~$\ge$~0.2\\
   			
   	\end{itemize}
   The limit on the variance of the second component (V(C$_2$)) was determined visually based on the first human classification. After manually classifying the systems into each of the four architecture classes, the visual distinction between what we considered \textit{Mixed} and what we considered \textit{Ordered} or \textit{Anti-Ordered} corresponds to a limit at 0.2. For a more detailed description of the choice of limits for max(M$_p$), S(C$_1$), and V(C$_2$), see Appendix \ref{App:bornes}\\
   Figure \ref{fig:syst_known} illustrates four examples of systems and their architecture. The Trappist-1 system, consisting of seven Earth-sized planets with masses ranging from 0.6 to 1.4~M$_{\oplus}$, is a prototypical example of a \textit{Low-mass} architecture \citep{Gillon2017}. In contrast, the Solar System is an \textit{Ordered} system \citep{Mishra2023}, characterised by its four small telluric planets followed by its four gaseous giants or icy planets, representing a distinct telluric/giant dichotomy. On the other hand, the architecture of HD 82943 is \textit{Anti-ordered}, featuring two Uranus-like planets followed by a small rocky planet. Finally, the architecture of GJ 433 is classified as \textit{Mixed}, with a super-Earth planet followed by one gaseous giant and a Neptune (see Fig. \ref{fig:syst_known}).
   \begin{figure}[h!]
   	\centering
        \includegraphics[width=8cm]{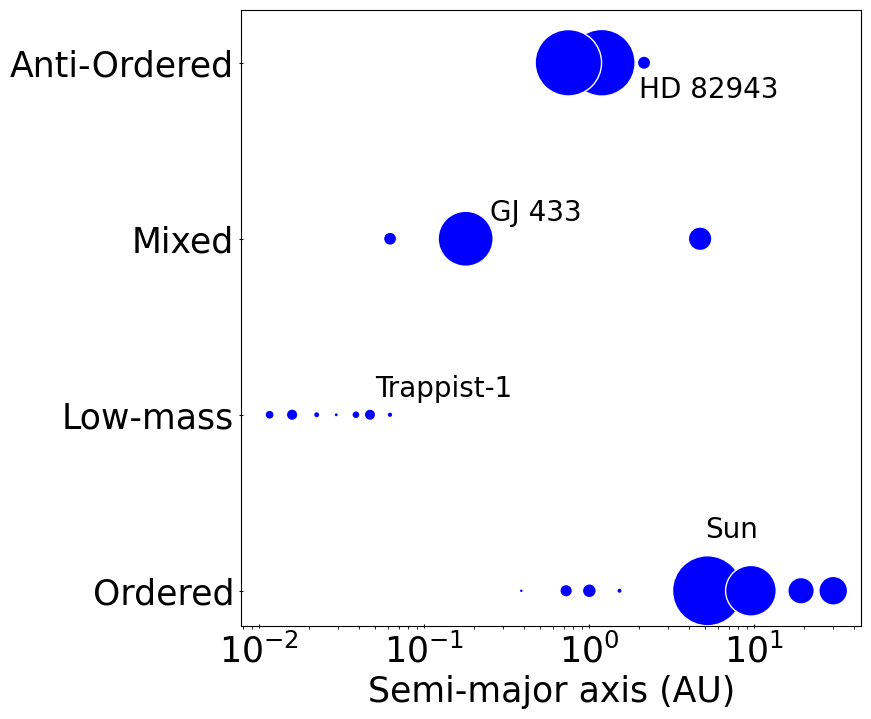}
        \caption{Examples of four different planetary systems (HD 82943, GJ 433, Trappist 1, and the Solar System) with different architectures. The size of the dots gives a representation of the planetary masses but not on the scale (the smallest have been multiplied up to 140 and the biggest divided by 2 to be able to represent all the planets on the same diagram).}
   	\label{fig:syst_known}
   \end{figure}
   
        \subsubsection{Theoretical and biased architectures} \label{subsubsection:theo_bias_archi}
\begin{figure}[h!]
   	\centering
   	\includegraphics[width=8cm]{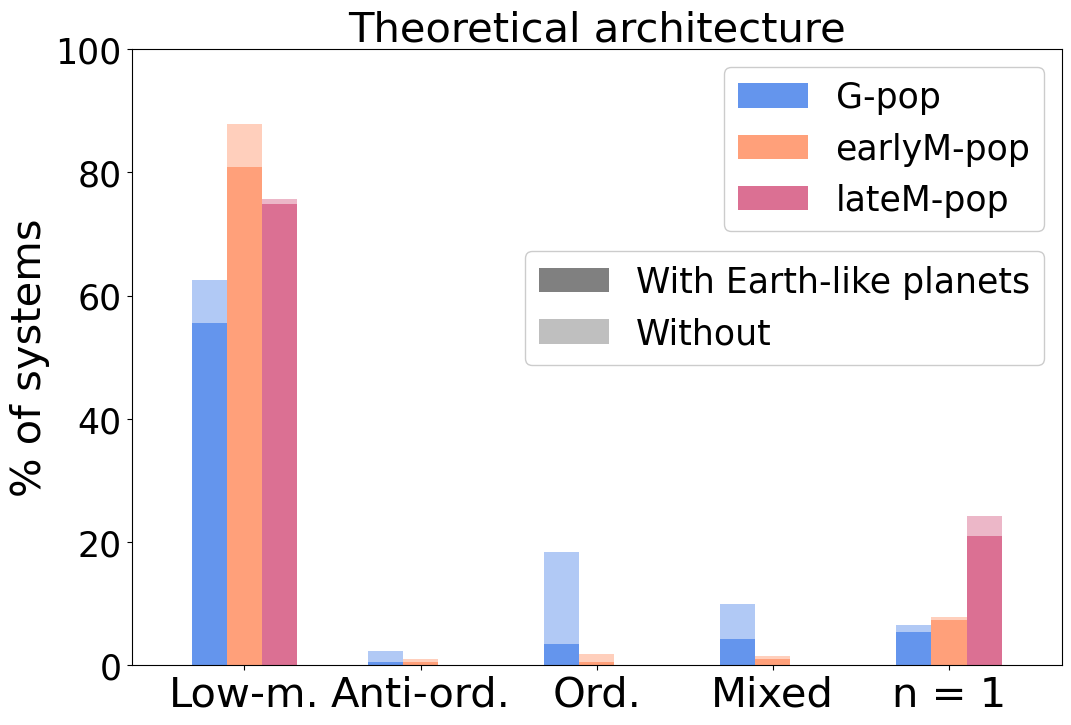}
        \includegraphics[width=8cm]{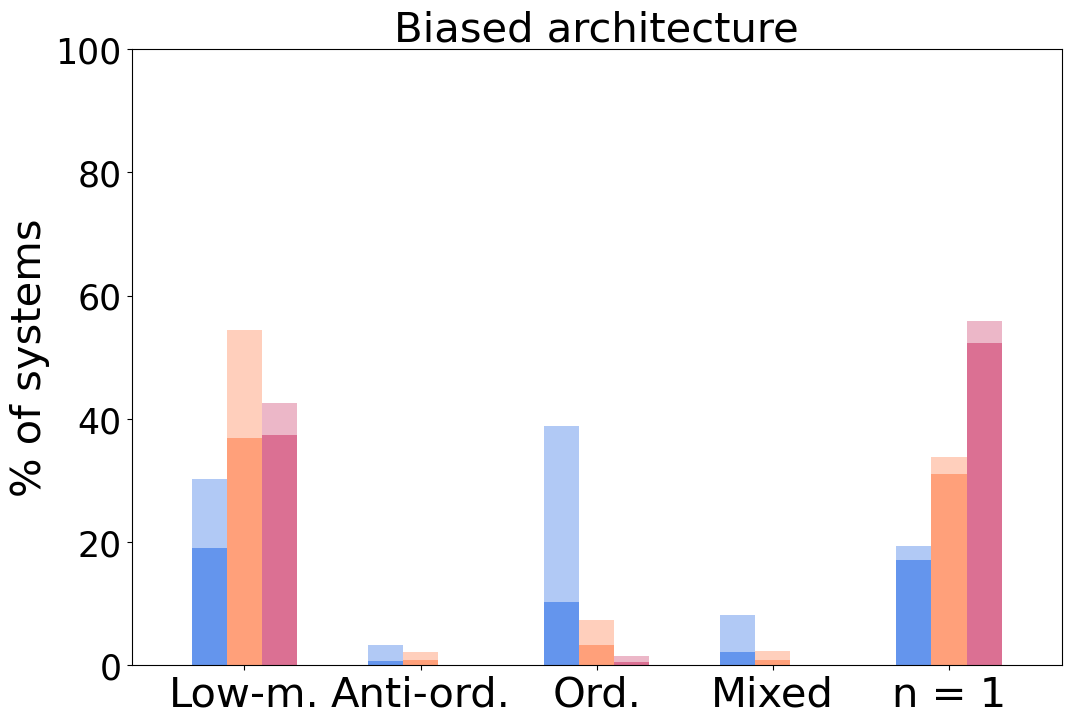}
        \caption{Distribution of the three populations in different architecture classes and proportion of systems with ELPs. The top panel shows theoretical architecture distribution, and the lower panel shows the distribution of biased architecture. The fully saturated colour bars represent the proportion of systems in each category with at least one ELP, and the greyed part represents the proportion of systems without ELP. There is, as an example, 63\% of \textit{Low-mass} systems around G stars, out of which 88\% host an ELP.}
   	\label{fig:distribution_archi_terre}
        \end{figure}

        When applying the architectural classification framework described beforehand on the three populations of synthetic planetary systems, we take into account another category of architecture: \textit{n~=~1}, which hosts all systems with a single planet.\\ 
        The three synthetic populations distributed in these five classes are represented in Fig. \ref{fig:distribution_archi_terre}, with (top panel) and without (lower panel) observational bias. The sum of the solid and shaded bars represents the percentage of a population (represented by colours) belonging to one of the five architectures. The solid-coloured portion alone corresponds to the fraction of such a category (e.g. \textit{Low-mass} systems of G-pop) that has at least one ELP. In contrast, the shaded portion represents the fraction of systems in the same category without ELP. The percentage of systems of each population in each theoretical architectural class is in table \ref{tab:perc_archi_avantbiais}, and in table \ref{tab:perc_archi_apresbiais} for the biased architecture classes. The corresponding percentage of systems with ELP is indicated in table \ref{tab:perc_ELP_avantbiais} for theoretical architectures and in table \ref{tab:perc_ELP_apresbiais} for biased architectures.\\
        The distribution of theoretical systems within the five architecture classes (top panel of Fig. \ref{fig:distribution_archi_terre}) exhibits a notable imbalance: between 63 and 88\% of synthetic systems among the three populations are \textit{Low-mass}, which means made of planets with small to moderate masses. EarlyM- and lateM-pop exhibit very few or no systems in the \textit{Anti-Ordered}, \textit{Ordered}, or \textit{Mixed} architectures. In almost all cases, these systems are either of \textit{Low-mass} architecture or consist of only one planet (\textit{n~=~1}). This can be attributed to the limited number of massive planets within these populations.
        On the other hand, G-pop exhibits 10\% to 18\% of systems in the \textit{Mixed} and \textit{Ordered} architectures, respectively, but only 2\% in the \textit{Anti-Ordered} architecture.
        This can be explained by theory: \textit{Anti-Ordered} architecture is a structure inducing a massive planet (M$_p$ > 20~M$_{\oplus}$) followed by smaller planets, which is an unlikely configuration for in-situ formation. Indeed, the presence of a massive or giant planet close to the star is often associated with a planetary migration episode. Such a planet forms in colder regions with more solid material and migrates to warmer regions closer to the star. These migration episodes could then, in many cases, disrupt the orbits of rocky planets that formed in the hot, close regions to the star, leading to their ejection from the system or collisions resulting in their destruction.\\
        The proportion of systems in the class \textit{n~=~1} increases when the mass of the central star decreases, which is explained by the fact that the planets of the populations of low stellar masses (0.2~M$_{\odot}$ in particular) are made up of planets of low masses, and therefore that they are the most sensitive to the fact of removing planets that are too small (M$_p$ < 0.5~M$_{\oplus}$): many systems end up having only one remaining planet.\\
        The distribution of systems hosting an ELP in these categories is also unequal. According to Table \ref{tab:perc_ELP_avantbiais}, most \textit{Low-mass} systems contain an ELP. Specifically, 88\% of \textit{Low-mass} systems in G-pop, 92\% in earlyM-pop, and 99\% in lateM-pop have an ELP. However, only between 18\% and 43\% of systems in G-pop with an \textit{Anti-Ordered}, \textit{Ordered}, or \textit{Mixed} architecture have an ELP, representing a minority in each case. Systems with a single planet also host an ELP in a majority (81\% for G-pop).
        This finding implies that the presence of an ELP is correlated with the properties of the planets in the same system. ELPs tend to form in systems with only relatively small planets, and the discovery of a planet with a mass similar to Earth could lead to the discovery of similar planets at different periods in the same system. Therefore, information on the architecture and on other planets of a system is crucial for inferring the types of planets that may be present in the currently unobservable part of the system.   

	\begin{table}[h]
	\caption[]{Percentage of systems in each theoretical architecture. For example, 63\% of G-pop systems have a \textit{Low-mass} architecture.}
	\begin{tabular}{lccc}
		\hline
		& G-pop & earlyM-pop & lateM-pop \\
		\hline
		Low-mass & 63 & 88 & 76 \\
		Anti-Ordered & 2 & 1 & 0 \\
		Ordered & 18 & 2 & 0 \\
		Mixed & 10 & 2 & 0 \\
		n~=~1 & 7 & 7.8 & 24 \\
		\hline
		\label{tab:perc_archi_avantbiais}
	\end{tabular}
	\end{table}
 
	\begin{table}[h]
	\caption[]{Percentage of systems in each theoretical architecture class with an ELP. For example, 88\% of \textit{Low-mass} G-pop systems have an ELP.}
	\begin{tabular}{lccc}
		\hline
		& G-pop & earlyM-pop & lateM-pop \\
		\hline
		Low-mass & 88 & 92 & 99\\
		Anti-Ordered & 23 & 51 & N.A.\\
		Ordered & 18 & 29 & 0\\
		Mixed & 43 & 61 & 100 \\
		n~=~1 & 81 & 94 &  86\\
		\hline
		\label{tab:perc_ELP_avantbiais}
	\end{tabular}
   	\tablefoot{N.A. stands for non-applicable and corresponds to the case where there is no system in the category.}
	\end{table}

 \begin{table}[h]
	\caption[]{Percentage of systems in each biased architecture class. For example, 30\% of biased G-pop systems have a \textit{Low-mass} architecture.}
	\begin{tabular}{lccc}
		\hline
		& G-pop & earlyM-pop & lateM-pop \\
		\hline
		Low-mass & 30 & 54 & 42 \\
		Anti-Ordered & 2 & 1 & 0 \\
		Ordered & 39 & 7 & 2 \\
		Mixed & 8 & 2 & 0 \\
		n~=~1 & 19 & 34 & 56 \\
		\hline
		\label{tab:perc_archi_apresbiais}
	\end{tabular}
\end{table}

\begin{figure}[h]
    \centering
    \includegraphics[width=6cm]{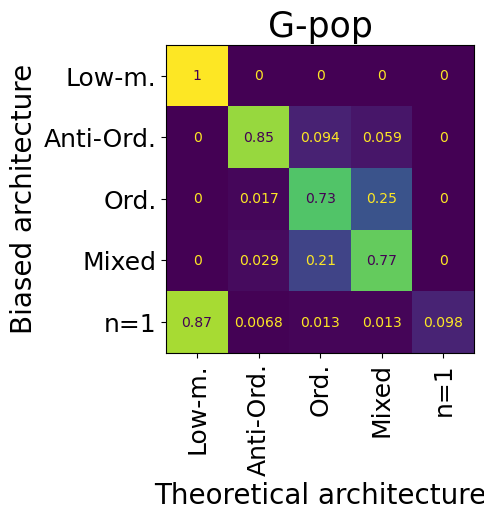}
    \includegraphics[width=6cm]{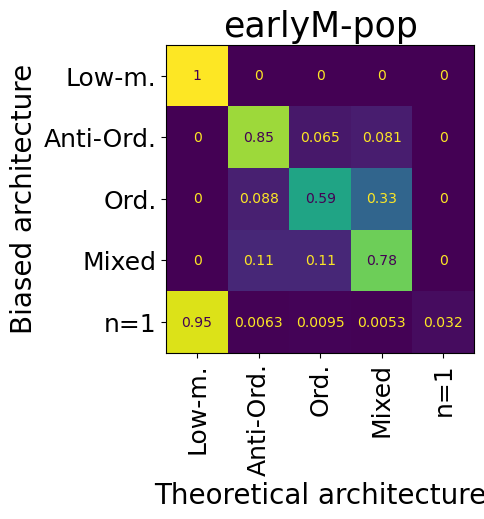}
    \includegraphics[width=6cm]{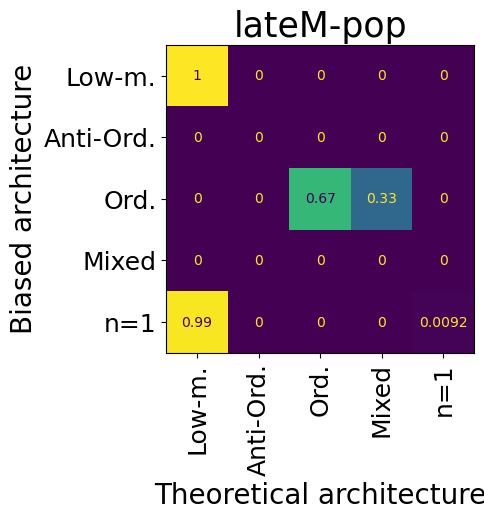}
    \caption{Confusion matrix representing the class change of each population with observational bias. The upper, middle, and lower panels represent the G-pop, earlyM-pop, and lateM-pop, respectively. The matrix is normalised on the biased architecture (lines). As an example, 85\% of biased \textit{Anti-Ordered} systems around G stars have an \textit{Anti-Ordered theoretical} architecture.}
    \label{fig:conf_matr}
    \end{figure}

     \begin{table}[h]
        \caption[]{Percentage of systems in each biased architecture class with ELP. For example, 63\% of biased \textit{Low-mass} G-pop systems have an ELP.}
	\begin{tabular}{lccc}
		\hline
		& G-pop & earlyM-pop & lateM-pop \\
		\hline
		Low-mass   & 63 & 68 & 88\\
		Anti-Ordered & 21 & 42 & N.A.\\
		Ordered & 26 & 45 & 33\\
		Mixed & 27 & 38 & N.A. \\
		n~=~1 & 88 & 92 & 94\\
		\hline
		\label{tab:perc_ELP_apresbiais}
	\end{tabular}
    \end{table}
    However, the observation bias selects only a few planets in a system that define the new, biased architecture. The distribution of biased synthetic planetary systems in the five architecture classes is shown in the lower panel of Fig. \ref{fig:distribution_archi_terre}. These proportions have changed drastically compared to the upper panel. Only 63\% of biased \textit{Low-mass} system in G-pop and 68\% in earlyM-pop have an ELP (see table \ref{tab:perc_ELP_apresbiais}.\\
    To further understand the effect of the observation bias on the architecture of synthetic populations, Fig. \ref{fig:conf_matr} shows how the theoretical architecture classes have changed for each synthetic population when considering the bias. 
    The upper, middle, and lower panels correspond to the G-pop, earlyM-pop, and lateM-pop respectively. The confusion matrices are normalised over the biased architectures axis (lines).
    Focussing on the \textit{Low-mass}, \textit{Anti-Ordered}, \textit{Ordered}, and \textit{Mixed} architectures, it is interesting to observe that the highest proportion of systems was already in those architectures without considering the bias, for G-pop and earlyM-pop.
    Only the class \textit{n~=~1} comprises mainly systems originally in the theoretical \textit{Low-mass} class, and only a few systems are initially systems with a single planet.
    In summary, it can be inferred that, unless for the biased class \textit{n~=~1}, the biased architecture of a system is representative of its theoretical architecture.

    \subsubsection{Comparison with observed systems}\label{biased_systems}
\begin{figure}[h!]
	\centering
	\includegraphics[width=8cm]{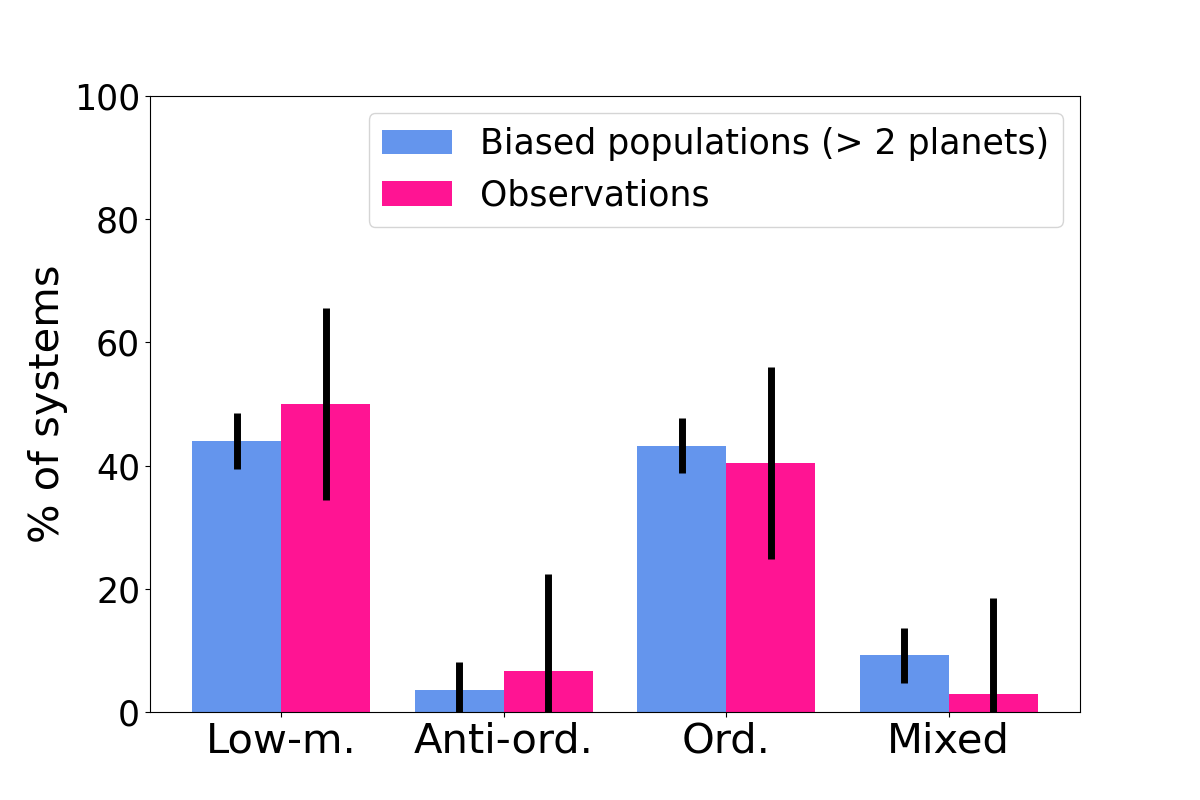}		
        \caption{Distribution of planetary systems in different classes of architecture. Blue bars represent the three biased synthetic populations, and the pink bars represent observed systems. Only the systems with at least three planets remaining after applying the bias are kept for the synthetic populations. The observed systems correspond to systems with at least three known planetary masses (or M$_p\cdot$ sin$i$, with $i$ the inclination of the system), and the same RV semi-amplitude threshold has been applied. The error bars correspond to 1/$\sqrt{N}$.}
	\label{comparaison_3p+obs}
        \end{figure}

        We can verify the relevance of the synthetic populations at the system level with the applied observation bias. This is done by comparing the biased architecture of the synthetic systems to the architecture of observed multiplanet systems. Therefore, a sample of 116 observed systems\footnote{The complete list is available here: \url{https://github.com/jdavoult/Architectures\_syst)}} comprising three or more planets with known mass (or M.sin(i)) is used. We apply the same RV bias to each observed system as to the synthetic populations based on the host star's mass. We only consider systems that retain at least three planets and we keep the same proportion of G, early-M, and late-M stars in both the observed and the biased population.\\
        Figure \ref{comparaison_3p+obs} compares the distribution of systems in the four different architecture classes between the synthetic systems from the three biased populations (all together, the blue bars) with at least three planets remaining after applying the bias and the observed systems mentioned beforehand. Overall, the distribution of synthetic system architecture classes replicates the distribution of observed systems: the \textit{Low-mass} and \textit{Ordered} architectures are the dominant classes, accounting for 42 and 45\%, respectively, for biased synthetic systems, and 58 and 36\% for observed systems. On the other hand, the \textit{Anti-Ordered} and \textit{Mixed} architecture categories host the least systems, comprising 3 and 11\% for biased synthetic systems and 2 and 4\% for observations. This makes it possible to consider the biased synthetic populations of systems as analogous to the observed systems.\\

        The theoretical architecture of a system gives us important information on the presence of an ELP within it. In particular, we find a large proportion of theoretical systems \textit{Low-mass} and \textit{n~=~1} with at least one ELP and a more variable proportion in the other three architecture classes. We have seen that the observation bias alters the architecture classes. However, a system's biased architecture remains a relevant element in this study and, in most cases, reflects its theoretical architecture. The comparison between the proportion of biased and observed systems in each architecture confirms the relevance of the synthetic populations and the observation bias chosen for this study.
    \subsection{Properties of the innermost detectable planet}
    As we have seen, ELP presence correlates with the properties of other planets in their systems and their arrangement. In particular, there is a correlation between the presence of an ELP and the properties of the innermost planet, which either formed in the inner region of the protoplanetary disc or formed in cooler regions and migrated inwards. Thus, its properties, such as its mass, radius, or period, can provide information on the conditions of planet formation --- its mass and radius can serve as clues to the mass and composition of the disc --- or on the planets themselves. Notably, the period often indicates whether the planets are grouped close to or far from the star. Especially for Peas-in-a-pod systems (comprising only small to medium planets), inner planets are often packed and/or in resonance within a few au. Therefore, if the innermost planet is found very close to the star, the rest of the system is more likely to be located very close to the star, leaving the temperate zone around the star empty. Conversely, if the first observed planet is relatively distant from the star, the system is shifted outwards, increasing the probability of finding a planet within the temperate zone around the star. Specific to the Bern model, this correlation might also depend on the initial number of embryos in the disc, governing the final number of planets in the system. \\
    The innermost detectable planet (IDP) is defined as the innermost planet that overcomes observational biases, and its properties are studied along with the architecture of the system.\\

    The biased architecture and properties of the IDP provide substantial information about the theoretical system. Together, they give hints on the system's formation pathway and provide significant insights into the entire system, including the potential presence of an ELP. Then, the biased architecture of a system, along with the mass, radius, and period of its IDP, serve as descriptive features of a system in the following section for studying correlations within systems. Thus, in the following section, we study the conditional probabilities of the presence of an ELP jointly with the architecture and properties of the IDPs in the synthetic systems of the three populations.

\section{Correlations between the presence of an ELP and the properties of their systems} \label{sec:correlations}
    
    Here we investigate the interrelation between the architecture of a system, the properties of its IDP, and the presence of an ELP. All the results are summarised in the table \ref{tab:resume}.

\begin{figure*}[]
    \caption[]{Histograms of distribution of IDPs' properties among systems in G-pop for different biased architectures. The first column depicts the repartition of IDPs' masses (logarithm scale), the second column shows the repartition of IDPs' radii, and the third the repartition of IDPs' periods (logarithm scale). Each line corresponds to a different group of biased architecture among the G-pop. The first line shows the IDPs' properties in the \textit{Low-mass} systems in G-pop, the second line the \textit{Anti-Ordered} systems, third line is for the \textit{Ordered} systems, fourth for \textit{Mixed} systems, and fifth for the systems with only one visible planet (\textit{n~=~1}, in this case IDP refers to the only visible planet). The blues bars refer to systems with one ELP or more, and the orange bars refer to systems without ELPs.}
        \begin{tabular}{>{\centering\arraybackslash}m{3cm}|ccc}
        \multirow{2}{*}{} & \thead{Log of the mass of the IDP} & \thead{Radius of the IDP} & \thead{Log of the period of the IDP} \\
        & & & \\
        \hline
        G-pop:Low-mass\vspace{1cm} & (a) \includegraphics[width=3.4cm]{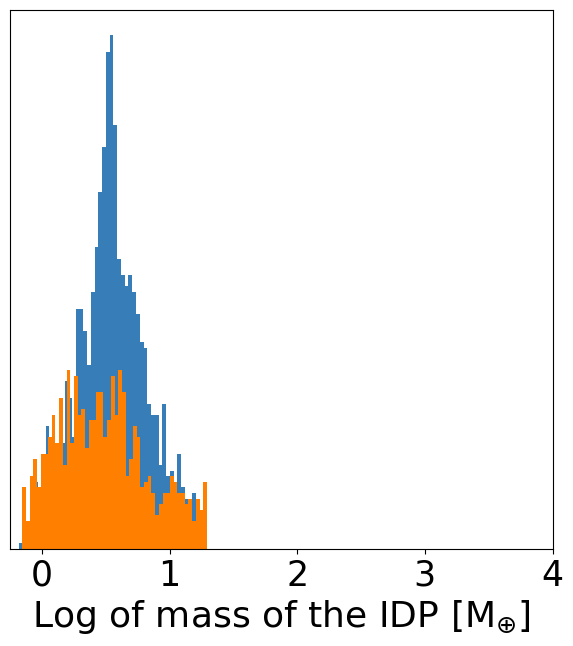} & (b) \includegraphics[width=3.4cm]{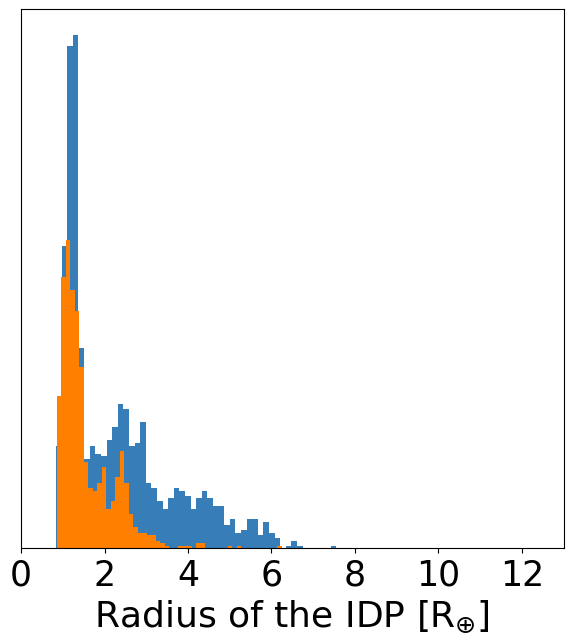} & (c) \includegraphics[width=3.4cm]{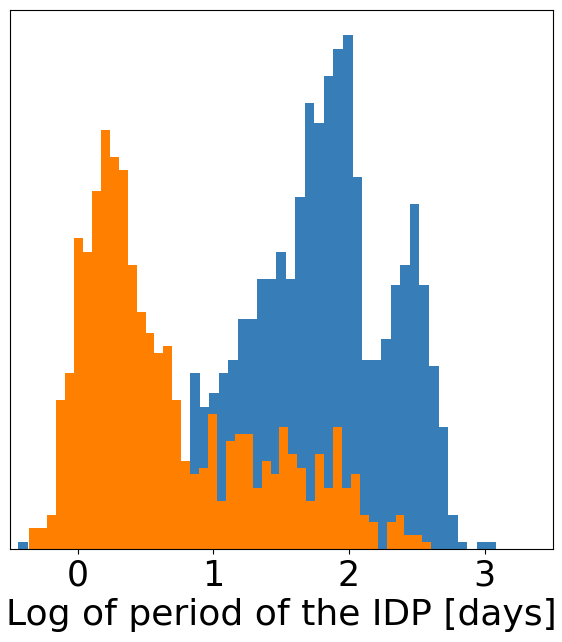}\\
        G-pop:Anti-Ordered\vspace{1cm} & (d) \includegraphics[width=3.4cm]{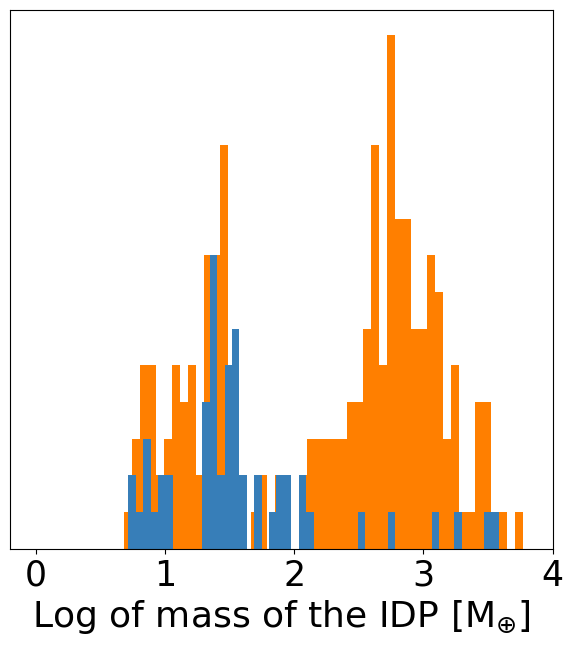} & (e) \includegraphics[width=3.4cm]{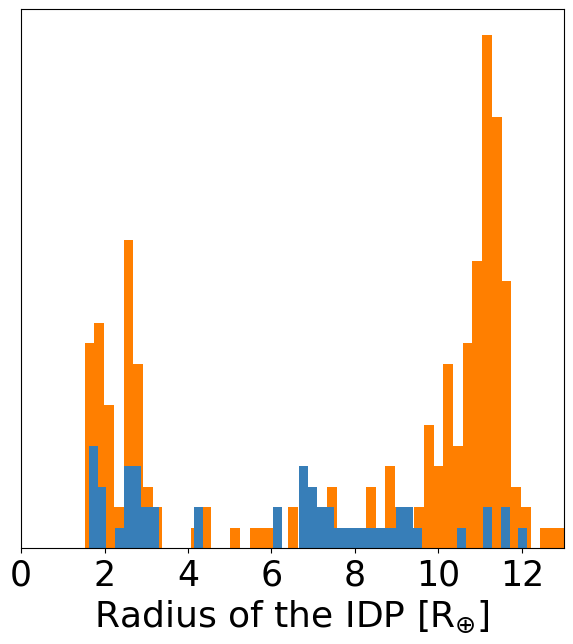} & (f) \includegraphics[width=3.4cm]{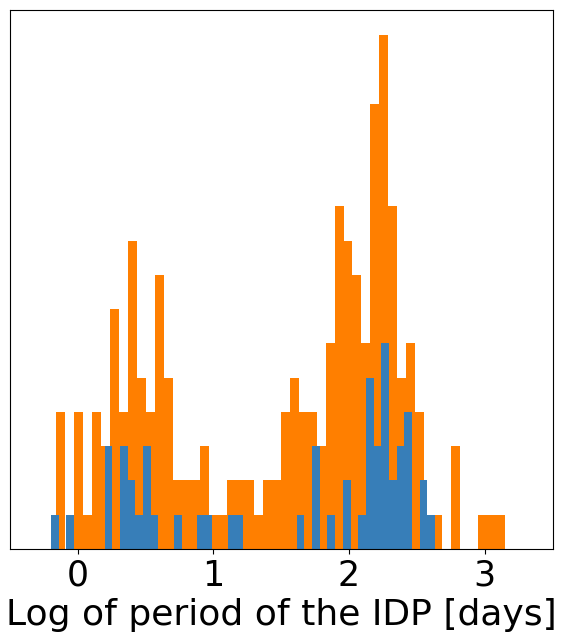}\\
        G-pop:Ordered\vspace{1cm} & (g) \includegraphics[width=3.4cm]{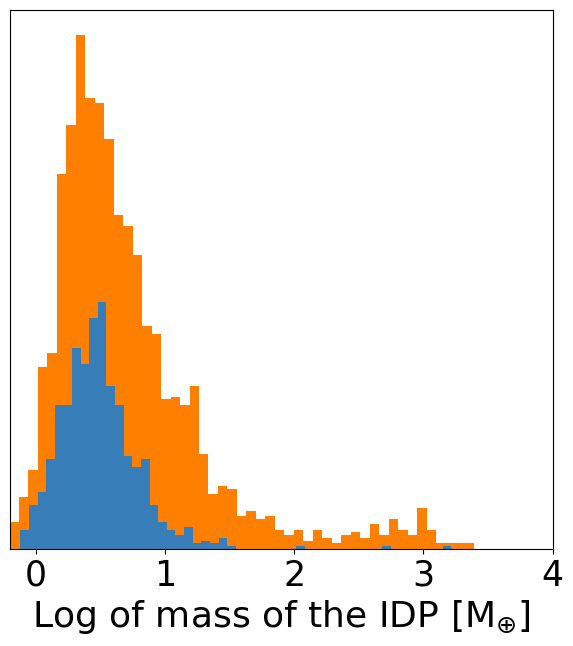} & (h) \includegraphics[width=3.4cm]{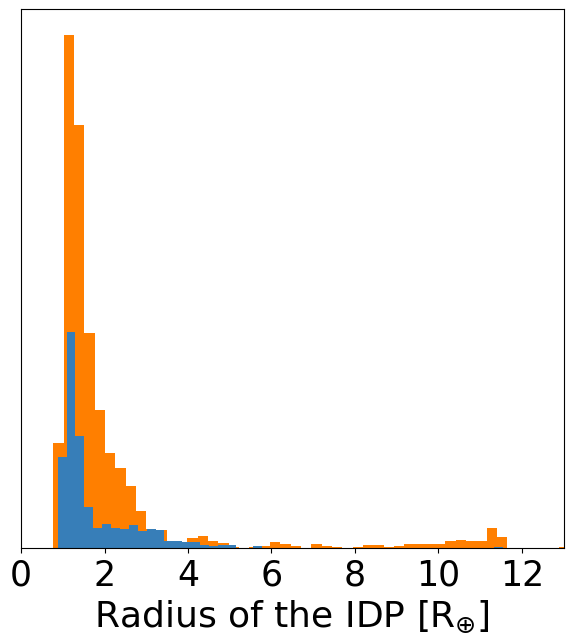} & (i) \includegraphics[width=3.4cm]{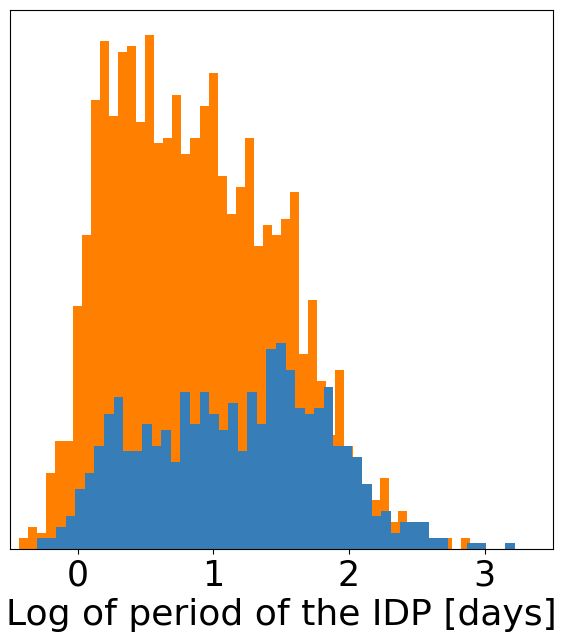}\\
        G-pop:Mixed\vspace{1cm} & (j) \includegraphics[width=3.4cm]{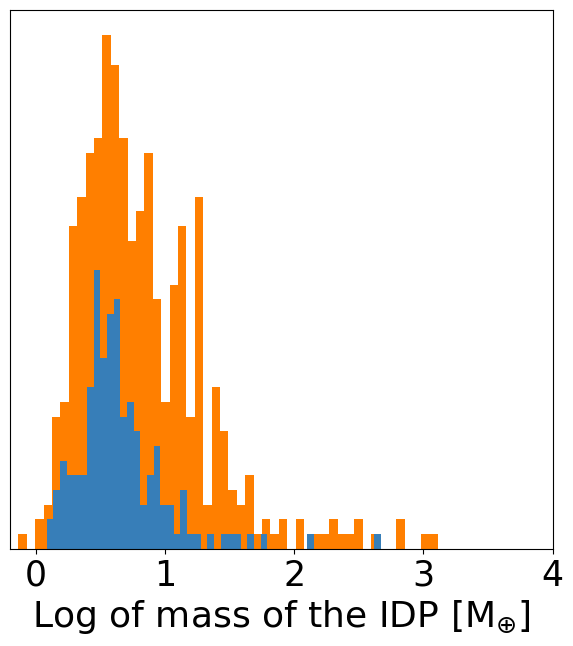} & (k) \includegraphics[width=3.4cm]{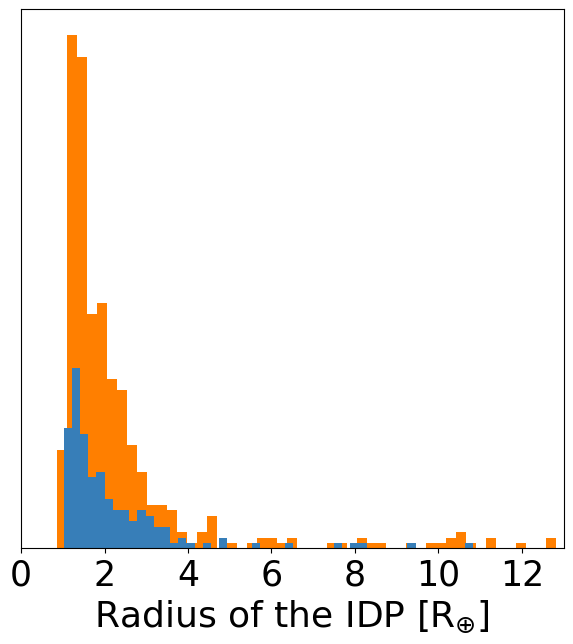} & (l) \includegraphics[width=3.4cm]{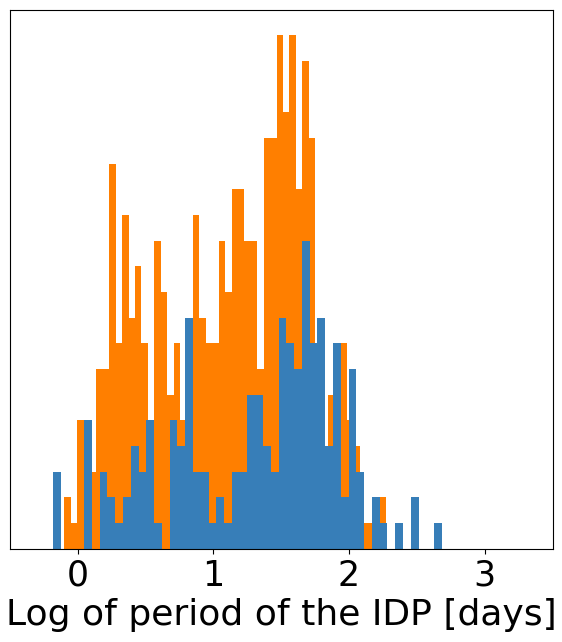}\\
        G-pop:n~=~1\vspace{1cm} & (m) \includegraphics[width=3.4cm]{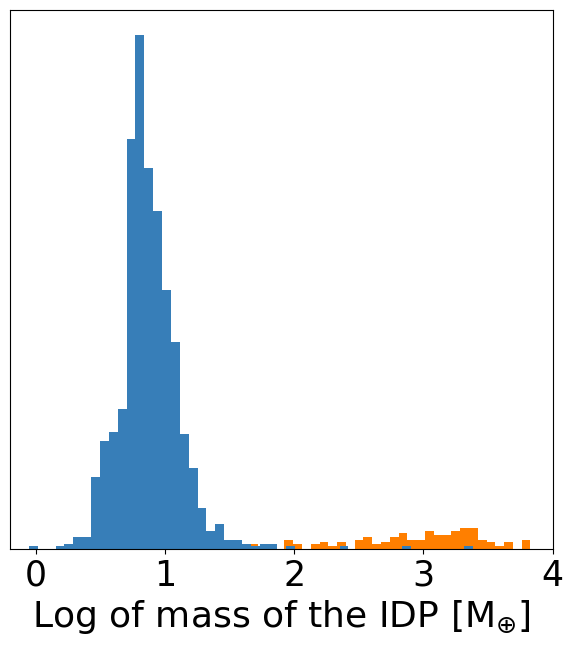} & (n) \includegraphics[width=3.4cm]{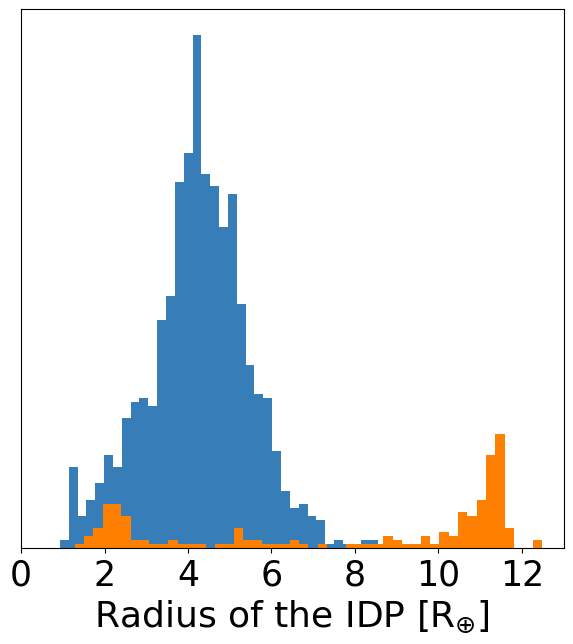} & (o) \includegraphics[width=3.4cm]{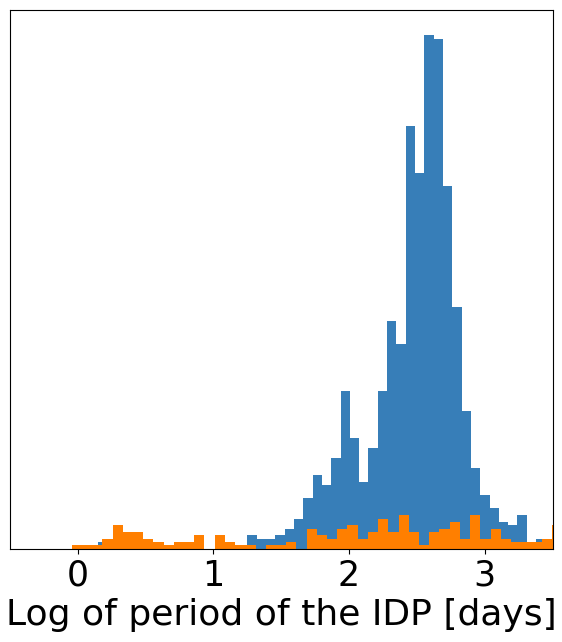}\\
    \end{tabular}
    \label{fig:resultsG}
    \end{figure*}

    \paragraph{\textbf{G-pop: M$_{\star}$~=~1~M$_{\odot}$}}
    Fig. \ref{fig:resultsG} presents histograms of the distributions of IDP properties for each category in G-pop. The columns represent different properties, and the rows represent different system architectures. The rows correspond to the architectures \textit{Low-mass}, \textit{Anti-Ordered}, \textit{Ordered}, \textit{Mixed}, and \textit{n~=~1}, respectively. Blue corresponds to systems hosting an ELP, while orange represents ELP-free systems.\\
    Across the different architectures, we find trends in the distribution of IDPs' properties. When we look at the IDP mass distribution (first column of Fig. \ref{fig:resultsG}), we find a tendency to find only ELP-free systems (orange bars) above an IDP mass of 100~M$_{\oplus}$, except for the \textit{Low-mass} architecture. For the \textit{Ordered} and \textit{Mixed} systems, this distinction is even at 10~M$_{\oplus}$. This is intuitive for \textit{Ordered} systems since they tend to have increasing masses. Although it is possible to find smaller planets with larger periods, in which case V(C$_2$) increases, the criterion on V(C$_2$) prevents the system from having too many variations in too small a period range. We therefore have a very low probability of these systems hosting an intermediate-period Earth-mass planet in the temperate zone of the systems. In terms of conditional probabilities for the mass of the IDP, we find:
    \begin{align*}
        &P(ELP|G star \cap Anti-ord. \cap M_{IDP} > 100~M_{\oplus})=6\%\\
        &P(ELP|G star \cap Ord. \cap M_{IDP} > 10~M_{\oplus})=7.1\%\\
        &P(ELP|G star \cap Mixed \cap M_{IDP} > 10~M_{\oplus})=13.1\%\\
        &P(ELP|G star \cap n = 1 \cap M_{IDP} > 100~M_{\oplus})=3.6\%.
    \end{align*}
    These findings clearly show the tendency for systems to be ELP-free when the IDP is a relatively massive planet. Indeed, the formation or migration of a giant planet (M$_p$ > 100~M$_{\oplus}$) in the inner part of a system prevents the formation or the stability of a terrestrial planet on temperate orbits. Even if the planet is not a giant but only a few dozen Earth masses, its formation may, in many cases, have prevented the formation of other planets with similar orbits.
    For the architecture \textit{n = 1}, this very clear result can also be explained by the architecture of the system. In the confusion matrix of figure \ref{fig:conf_matr} for G-pop we see that biased \textit{n~=~1} systems have mostly theoretical \textit{n~=~1} or \textit{Low-mass} architectures (95\% of them). If the IDP is more massive than 20~M$_{\oplus}$ they cannot be \textit{Low-mass}, and if they are theoretically \textit{n~=~1}, then the IDP is the only planet in this system, which explains the quasi-absence of ELPs in these systems. The rest of the systems with smaller IDPs are therefore either theoretically \textit{Low-mass} (87\%), which ensures a very high probability of hosting an ELP, or \textit{n~=~1} (10\%), which are the few systems not to host an ELP.\\
    
    As for the radius of the IDP (second column of Fig. \ref{fig:resultsG}), for the architectures \textit{Anti-Ordered}, \textit{Ordered}, \textit{Mixed}, and \textit{n = 1}, we see a tendency for the systems to be ELP-free when the radius is large. The limit can be drawn at different values for each architecture, but the trend remains the same and matches the result found for the mass. For the \textit{Low-mass} systems, however, we find the opposite: the systems tend to be ELP-free mostly when the radius is lower, which indicates a tendency for IDP to be less dense in ELP-hosting systems.
    In terms of conditional probabilities, we find:
    \begin{align*}
        &P(ELP|G stars \cap Low-m. \cap R_{IDP} > 2.5~R_{\oplus}) = 87.5\% \\
        &P(ELP|G stars \cap Anti-Ord. \cap R_{IDP} > 10~R_{\oplus}) = 5.3\% \\
        &P(ELP|G stars \cap Ord. \cap R_{IDP} > 6~R_{\oplus}) = 2.9\% \\
        &P(ELP|G stars \cap Mixed \cap R_{IDP} > 2.5~R_{\oplus}) = 8.33\% \\
        &P(ELP|G stars \cap n = 1 \cap R_{IDP} > 8~R_{\oplus}) = 7.6\%. 
    \end{align*}

    For the period (third column of Fig. \ref{fig:resultsG}), we see a general tendency for systems to be mostly ELP-free when the IDP period is too small (except for systems \textit{Anti-Ordered}, where we do not notice a particular tendency on panel (f)). This limit is around 30~days more or less for all architectures, except \textit{Low-mass} for which it is around 10~days. In terms of probability, we find:
    \begin{align*}
        &P(ELP|G stars \cap Low-m. \cap P_{IDP} > 10~days) = 79.2\% \\
        &P(ELP|G stars \cap Ord. \cap P_{IDP} < 30~days) = 14.1\% \\
        &P(ELP|G stars \cap Mixed \cap P_{IDP} < 30~days) = 13.8\% \\
        &P(ELP|G stars \cap n = 1 \cap P_{IDP} < 30~R_{\oplus}) = 37\%. \\
    \end{align*}
   This can be explained by the tendency of planets to gather on close orbits, as was mentioned earlier. Therefore, if the IDP is in the temperate zone or very close to it, one of the following planets is very likely to be in the temperate zone. Conversely, if the IDP is very close to the star (P~<~30~days), the system is shifted inwards and the probability of finding a planet in the temperate zone drops. However, a planet can be found in the temperate zone if the ‘inner chain of planets’ is long (typically more than six planets in our simulations).

\begin{figure*}[h]
	\begin{tabular}{>{\centering\arraybackslash}m{3cm}|ccc}
        \multirow{2}{*}{} & \thead{Log of the mass of the IDP} & \thead{Radius of the IDP} & \thead{Log of the period of the IDP} \\
        & & & \\
        \hline
            earlyM-pop:Low-mass\vspace{1cm} & (a) \includegraphics[width=3.4cm]{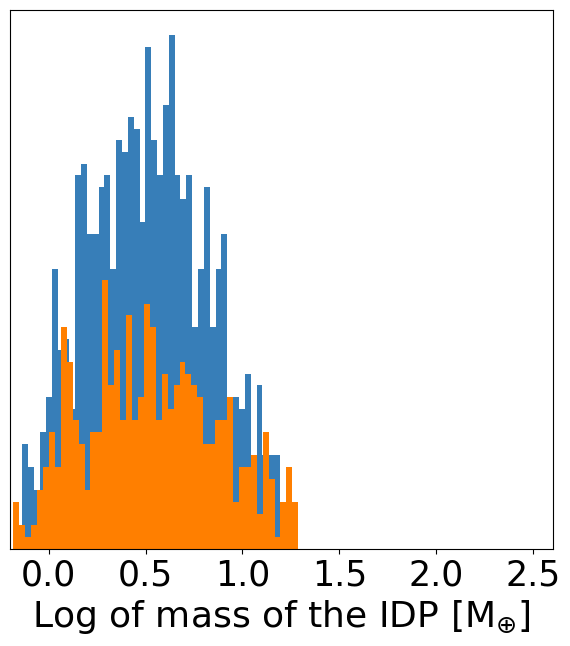} & (b) \includegraphics[width=3.4cm]{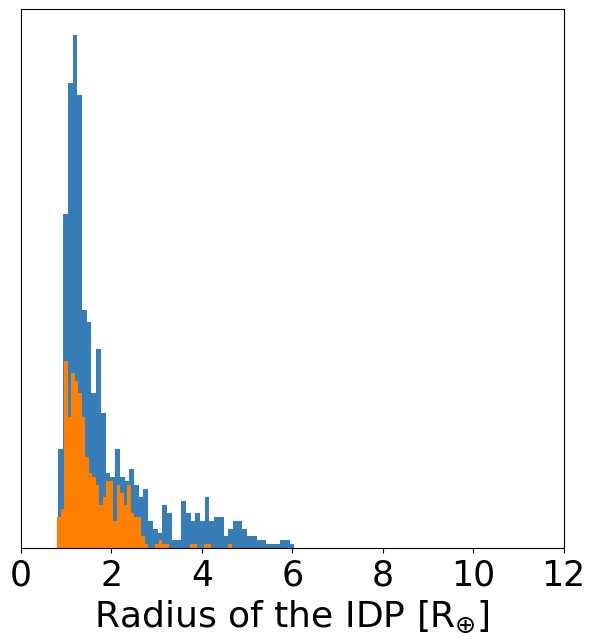} & (c) \includegraphics[width=3.4cm]{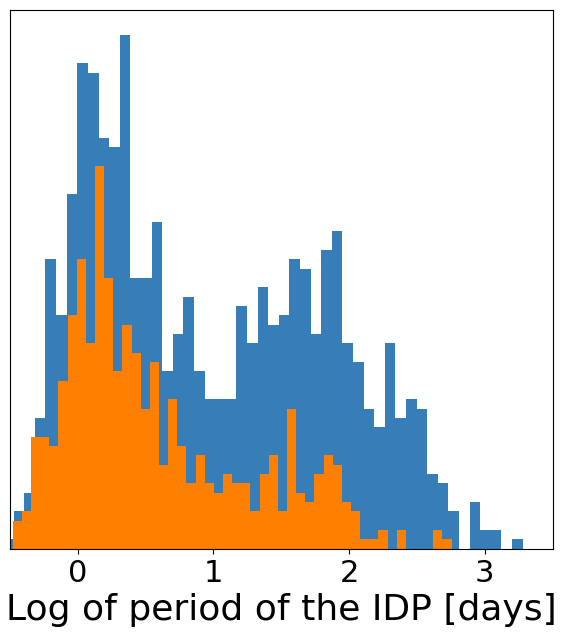} \\
            earlyM-pop:Ordered\vspace{1cm} & (a) \includegraphics[width=3.4cm]{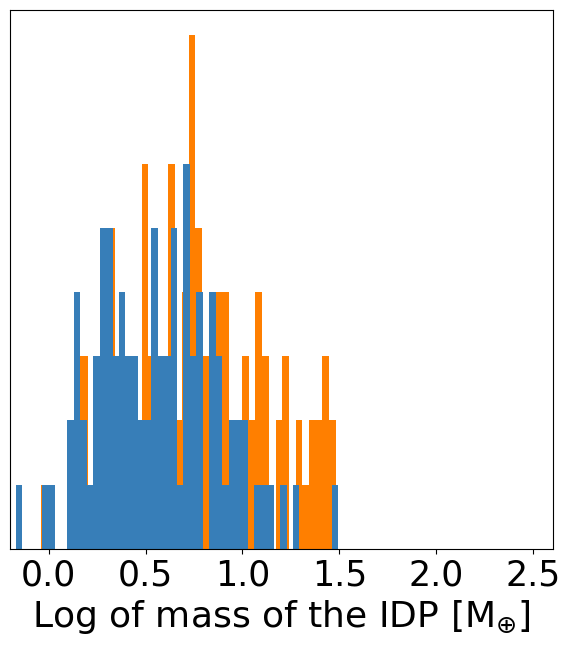} & (b) \includegraphics[width=3.4cm]{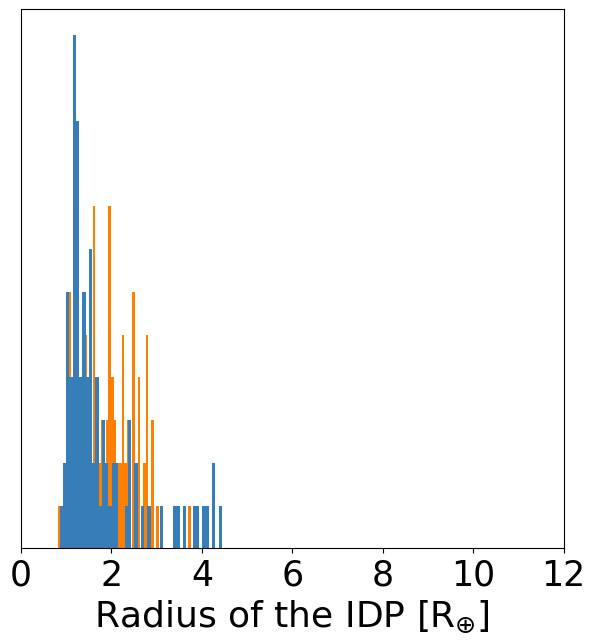} & (c) \includegraphics[width=3.4cm]{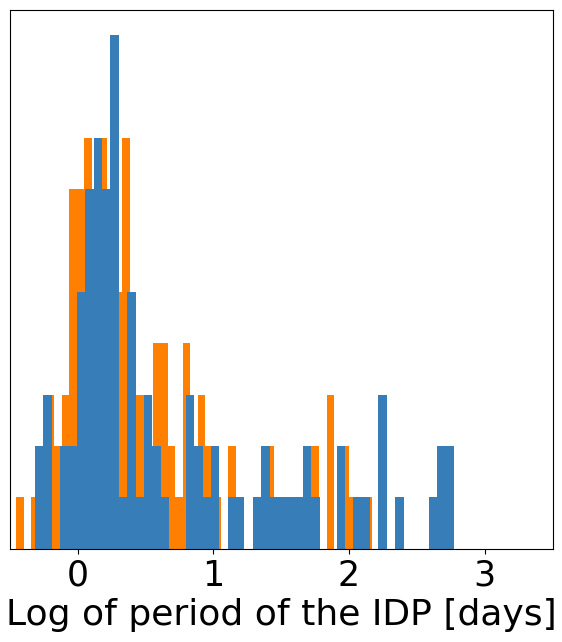} \\
            earlyM-pop:n~=~1\vspace{1cm} & (d) \includegraphics[width=3.4cm]{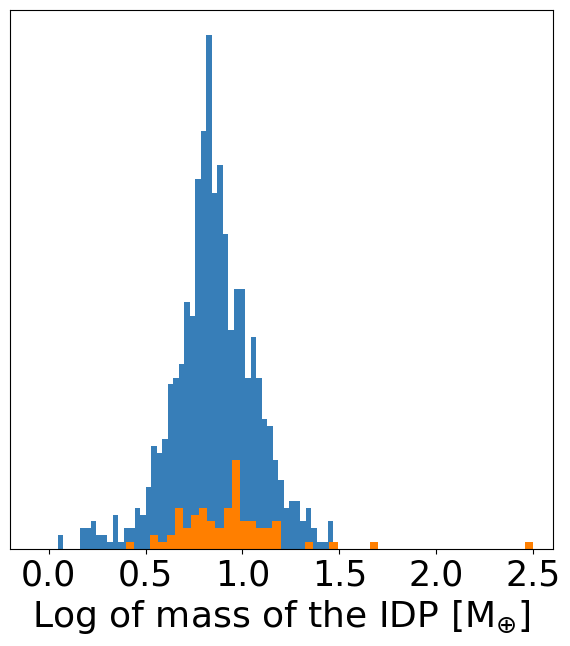} &(e) \includegraphics[width=3.4cm]{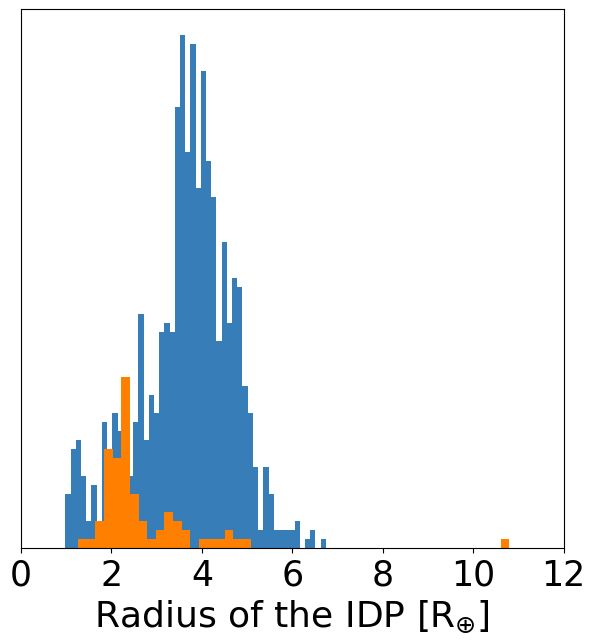} & (f) \includegraphics[width=3.4cm]{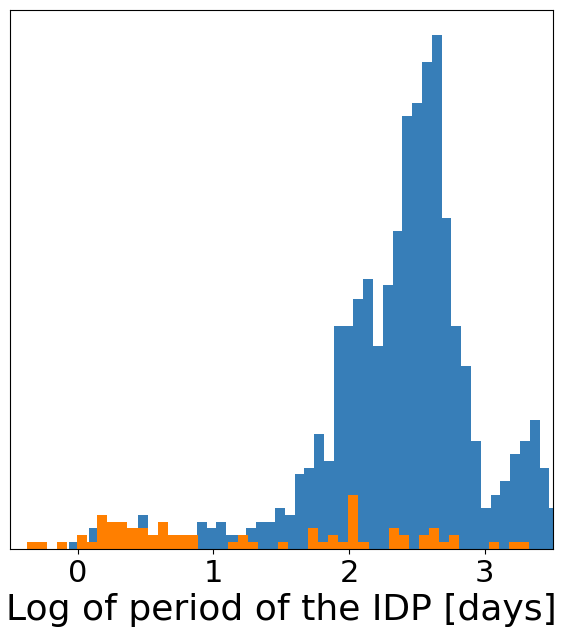} \\
		
	\end{tabular}
    \caption[]{Figure equivalent to Fig. \ref{fig:resultsG} for EarlyM-pop. The first, second, and third columns depict the repartition of IDPs' masses (logarithm scale), radii, and periods (logarithm scale). Here, only the systems \textit{Low-mass} (first line) and \textit{n~=~1} (second line) systems are shown because the three other groups do not give informative insights. The blues bars refer to systems with one ELP or more, and the orange bars refer to systems without ELPs.}\label{fig:resultsK}
    \end{figure*}

    \paragraph{\textbf{earlyM-pop: M$_{\star}$~=~0.5~M$_{\odot}$}}
    Fig. \ref{fig:resultsK} is equivalent to Fig. \ref{fig:resultsG}, it shows the distribution of mass (first column), radius (second column), and period (third column) of the IDPs for systems with a biased architecture \textit{Low-mass} (first row), \textit{Ordered} (second row), and \textit{n = 1} (third row), either with an ELP (blue) or ELP-free (orange). The trends are similar to those for the G-pop systems but differ slightly. The results for the \textit{Anti-ordered} and \textit{Mixed} systems are not shown because there are too few systems in each category (only 62 and 63, respectively).\\
    To begin with, the mass of the IDP does not give any information, either for the \textit{Low-mass} architecture or for the \textit{n = 1} architecture. There is a slight tendency for systems with slightly more massive IDPs to be ELP-free, as in G-pop. We find:
    \begin{align*}
        &P(ELP|early-M stars \cap Ord. \cap M_{IDP} > 10 M_{\oplus}) = 21.6\%
    \end{align*}
    In terms of the radius of the IDPs, we find the same tendency for \textit{Low-mass} systems in earlyM-pop than in G-pop to host an ELP when the radius of the IDP is greater than 2.75~R$_{\oplus}$. There is a very slight tendency for \textit{Ordered} systems to be ELP-free at a higher radius of IDPs, but unlike G-pop, \textit{n = 1} ELP-free systems rather tend to have an IDP with a low radius. We find:
    \begin{align*}
        &P(ELP|early-M stars \cap Low-m. \cap R_{IDP} > 2.75 R_{\oplus}) = 94.6\%\\
        &P(ELP|early-M stars \cap Ord. \cap R_{IDP} > 2 R_{\oplus}) = 33.8\%\\
        &P(ELP|early-M stars \cap n = 1 \cap R_{IDP} > 2.75 R_{\oplus}) = 97.3\%
    \end{align*}
    The third column shows the period distribution of the IDPs of systems with and without ELPs. In the same way as for G-pop, ELPs tend to be found in systems with IDP with larger periods. We can see this trend very clearly for the \textit{Low-mass} and \textit{n = 1} systems, less so for the \textit{Ordered} systems:
    \begin{align*}
        &P(ELP|early-M stars \cap Low-m. \cap P_{IDP} > 10~days) = 79.2\%\\
        &P(ELP|early-M stars \cap Ord. \cap P_{IDP} > 10~days) = 95.8\%.
    \end{align*}
    
    \paragraph{\textbf{lateM-pop: M$_{\star}$~=~0.2~M$_{\odot}$}}
    LateM-pop is a unique case in the study. The vast majority (more than 90\%) of systems with at least one visible planet harbour ELPs, indicating a high likelihood of finding ELPs in such systems. However, delving into the details, almost only the systems of \textit{Low-mass} and \textit{n~=~1} architectures have detectable planets, as the \textit{Anti-Ordered}, \textit{Ordered}, and \textit{Mixed} classes have only between zero and three systems each (in the Table \ref{tab:resume}, these categories will be denoted as N.A. for not applicable).\\
    \paragraph{\textbf{Summary of the results}}
    Table \ref{tab:resume} summarises the conditional probabilities of synthetic systems obtained through the detailed study of the properties of their IDPs, whether they host an ELP or not. The vertical entry depends on the type of central star, the horizontal entry depends on the type of architecture formed by the visible planets in the system, and the values in the table describe the conditional probabilities of a system hosting an ELP or not, based on the properties of its IDP.
    M$_{IDP}$, R$_{IDP}$, and P$_{IDP}$ refer to the mass, radius, and period respectively, of the IDP in biased systems in one of the four architectures, and of the only detectable planet in systems in categories \textit{n~=~1}. In the table, N.A. corresponds to categories with too few systems to draw conclusions. The percentages indicate the percentage of systems with an ELP in each case.
    As an example, the probability for a system orbiting a Sun-like star with a \textit{Low-mass} architecture, and an IDP radius greater than 2.5~M$_{\oplus}$ to have an ELP is:
    \begin{equation*}
        P(ELP|G star \cap Low-mass \cap R_{IDP} > 2.5~R_{\oplus}) = 87.5\%.
    \end{equation*}
    \begin{table*}[h!]
	\begin{center}
            \caption{Summary of the conditional probabilities of a system to host an ELP depending on its biased architecture and the properties of its IDP.}
		\begin{tabular}{l|c|c|c}
			& \textbf{G-pop (1~M$_{\odot}$)} & \textbf{earlyM-pop (0.5~M$_{\odot}$)} & \textbf{lateM-pop (0.2~M$_{\odot}$)} \\
			\hline
                        & M$_{IDP}$ < 10~M$_{\oplus}$ $\Rightarrow$ 64.0\% & M$_{IDP}$ < 10~M$_{\oplus}$ $\Rightarrow$ 68.5\% & \\
                        & M$_{IDP}$ > 10~M$_{\oplus}$ $\Rightarrow$ 49.7\% & M$_{IDP}$ > 10~M$_{\oplus}$ $\Rightarrow$ 59.2\% & \\
			          & R$_{IDP}$ < 2.5~R$_{\oplus}$ $\Rightarrow$ 55.8\% & R$_{IDP}$ < 2.75~R$_{\oplus}$ $\Rightarrow$ 64.1\% & \\
	 \textit{Low-mass} & \textbf{R$_{IDP}$ > 2.5~R$_{\oplus}$ $\Rightarrow$ 87.5\%} & \textbf{R$_{IDP}$ > 2.75~R$_{\oplus}$ $\Rightarrow$ 94.6\%} & 88\% \\
				      & P$_{IDP}$ < 10~days $\Rightarrow$ 38.0\% & P$_{IDP}$ < 10~days $\Rightarrow$ 60.3\% & \\
			          & \textbf{P$_{IDP}$ > 10~days $\Rightarrow$ 82.7\%} & P$_{IDP}$ > 10~days $\Rightarrow$ 79.2\% & \\
			\hline
			          & M$_{IDP}$ < 100~M$_{\oplus}$ $\Rightarrow$ 37.5\% & & \\
                        & M$_{IDP}$ > 100~M$_{\oplus}$ $\Rightarrow$ 6\% &  & \\
			          & R$_{IDP}$ < 10~R$_{\oplus}$ $\Rightarrow$ 34.3\% & & \\
  \textit{Anti-Ordered} & R$_{IDP}$ > 10~R$_{\oplus}$ $\Rightarrow$ 5.3\% & N.A. & N.A. \\
                        & P$_{IDP}$ < 30~days $\Rightarrow$ 23.3\% & & \\
                        & P$_{IDP}$ > 30~days $\Rightarrow$ 20.2\% & & \\
			\hline
				      & M$_{IDP}$ < 10~M$_{\oplus}$ $\Rightarrow$ 30.6\% & M$_{IDP}$ < 10~M$_{\oplus}$ $\Rightarrow$ 50.0\% & \\
			          & M$_{IDP}$ > 10~M$_{\oplus}$ $\Rightarrow$ 7.1\% & M$_{IDP}$ > 10~M$_{\oplus}$ $\Rightarrow$ 21.6\% &  \\
	                & R$_{IDP}$ < 6~R$_{\oplus}$ $\Rightarrow$ 27.8\% & R$_{IDP}$ < 2~R$_{\oplus}$ $\Rightarrow$ 50.4\% & N.A. \\
       \textit{Ordered} & R$_{IDP}$ > 6~R$_{\oplus}$ $\Rightarrow$ 2.9\% & R$_{IDP}$ > 2~R$_{\oplus}$ $\Rightarrow$ 33.8\% & \\
                        & P$_{IDP}$ < 30~days $\Rightarrow$ 14.1\% & P$_{IDP}$ < 10~days $\Rightarrow$ 40.7\% & \\
                        & P$_{IDP}$ > 30~days $\Rightarrow$ 67.5\% & P$_{IDP}$ > 10~days $\Rightarrow$ 60.5\% & \\
			\hline
	                & M$_{IDP}$ < 10~M$_{\oplus}$ $\Rightarrow$ 31.7\% &  &  \\
				      & M$_{IDP}$ > 10~M$_{\oplus}$ $\Rightarrow$ 13.1\% & & \\
                        & R$_{IDP}$ < 2.5~R$_{\oplus}$ $\Rightarrow$ 27.4\% & & \\
         \textit{Mixed} & R$_{IDP}$ > 2.5~R$_{\oplus}$ $\Rightarrow$ 8.33\% & N.A. & N.A. \\
                        & P$_{IDP}$ < 30~days $\Rightarrow$ 13.8\% & & \\
                        & P$_{IDP}$ > 30~days $\Rightarrow$ 49.1\% & & \\
			\hline
			          & \textbf{M$_{IDP}$ < 100~M$_{\oplus}$ $\Rightarrow$ 94.8\%} & \textbf{M$_{IDP}$ < 10~M$_{\oplus}$ $\Rightarrow$ 92.8\%} & \\
			          & M$_{IDP}$ > 100~M$_{\oplus}$ $\Rightarrow$ 3.6\% & M$_{IDP}$ > 10~M$_{\oplus}$ $\Rightarrow$ 90.2\% & \\
			          & \textbf{R$_{IDP}$ < 8~R$_{\oplus}$ $\Rightarrow$ 95\%} & R$_{IDP}$ < 2.75~R$_{\oplus}$ $\Rightarrow$ 75.1\% & \\
			\textit{n~=~1} & R$_{IDP}$ > 8~R$_{\oplus}$ $\Rightarrow$ 7.6\% & \textbf{R$_{IDP}$ > 2.75~R$_{\oplus}$ $\Rightarrow$ 97.3\%} & 94\% \\
			          & P$_{IDP}$ < 30~days $\Rightarrow$ 37.0\% & P$_{IDP}$ < 10~days $\Rightarrow$ 49.3\% &  \\
			          & \textbf{P$_{IDP}$ > 30~days $\Rightarrow$ 90.6\%} & \textbf{P$_{IDP}$ > 10~days $\Rightarrow$ 95.8\%} & \\

		\end{tabular}
	\end{center}
	\label{tab:resume}
	\tablefoot{M$_{IDP}$, R$_{IDP}$, and P$_{IDP}$ refer to the mass, radius, and period respectively of the IDP in the system. Percentages indicate the percentage of systems hosting an ELP. N.A. corresponds to categories with too few systems to conclude. The results in bold highlight the cases with the highest probability of finding an ELP.}
    \end{table*}

\section{Discussion and conclusion} \label{sec:conclu}
The correlations between the presence of an ELP and the properties of the detectable planets in Earth-hosting systems have been investigated through a study of conditional probabilities. Synthetic planetary systems from the Bern model orbiting three types of stars (1, 0.5, and 0.2 M$_{\odot}$) were used. We applied an observational bias using an RV semi-amplitude threshold to simulate their observed properties. This threshold varies depending the central star's mass. We also define five classes of architectures: \textit{Low-mass}, \textit{Anti-Ordered}, \textit{Ordered}, \textit{Mixed}, and \textit{n~=~1}. Each system with at least one planet was classified into one of these classes according to the mass of its most massive planet, the results of a principal component analysis and the number of planets it contains. We define the theoretical and biased architecture as the architecture without and with observation bias. In most cases, the biased architecture is a good proxy for the theoretical architecture. These biased synthetic systems have been compared with observed systems, and the distribution of architectures in these two groups is similar, highlighting the relevance of the Bern model in this study.\\
An initial analysis of the populations of synthetic systems showed that ELPs, as we define them in this paper, are common around the stars studied. They are found around 60\% of G-type stars, 74\% of early-M type stars, and 40\% of late-M type stars. A more detailed study of the conditional probabilities between the presence of an Earth and the properties of its system enabled us to identify which systems are most likely to host an ELP.\\
The architecture and properties of the IDP were highlighted as relevant quantities to be related to the presence of an ELP in a system, as they both reflect, but in different ways, the formation conditions of the system. A study of the interconnection between the architecture and the properties of the IDP is therefore carried out using conditional probability calculations. The results show that for the population of synthetic systems around G stars, we can predict the presence of an ELP in most cases, as we find very high (P>0.8) or very low (P<0.1) conditional probabilities in every architecture class. In particular, biased \textit{Low-mass} system with the radius of the IDP larger than 2.5~R$_{\oplus}$ or a period greater than 10~days, and \textit{n~=~1} systems with the visible planet with a mass smaller than 100~M$_{\oplus}$, radius greater than 8~R$_{\oplus}$ or a period greater than 30~days are the systems that show the most probability to host an ELP. In contrast, biased \textit{Anti-ordered}, \textit{Ordered}, \textit{Mixed}, and \textit{n = 1} systems with an IDP too massive or too large are the systems that show the least probability to host an ELP, and this can be explained by the inhibition of the formation of a terrestrial planet.\\
For the population around early-M stars (0.5~M$_{\odot}$), the same trends have been found as in G-pop. However, biased \textit{Anti-ordered} and \textit{Mixed} systems are not studied because too few biased systems fall into these categories.\\
Concerning the lateM-pop, if a system falls into the \textit{Low-mass} or \textit{n~=~1} category, it is highly likely to contain an ELP. Investigating the conditional probabilities in other architecture classes is impossible because only a few systems fall into these categories. \\
In general, these results show that ELPs are less commonly found in systems with large planets (M$_p$ > a few tens M$_{\oplus}$) or short period planets (P < 10~days). This implies that the systems of \textit{Low-mass} and \textit{n~=~1} architectures are favoured. We present the final results of the conditional probability study in Table \ref{tab:resume}. The systems most likely to host an ELP among the \textit{Low-mass} and \textit{n~=~1} architectures are in bold.\\

It is interesting to note that the solar system does not fall into the categories of systems most likely to host an ELP, although it does host two: Venus and the Earth. When we apply the observation bias used in section \ref{subsec:bias}, there are three visible planets: Jupiter, Saturn, and Neptune. Its biased architecture is \textit{Anti-Ordered} and its IDP is Jupiter. In contrast, the results on other known systems hosting an ELP confirm the results of our study. We extracted a list of 24 systems from Encyclopaedia of exoplanetary systems\footnote{Available at \url{https://exoplanet.eu/home/}} that host at least one planet meeting our criteria in terms of mass and equilibrium temperature to be considered as an ELP. We then applied the same observation bias used in this study depending on the mass of the central star, which removed a few planets (including ELPs). After defining their architecture using the framework described in section \ref{subsubsec:framework} and identifying their IDP, it appears that 100\% of these systems fall into one of the categories identified as likely to host an ELP. This unequivocal result confirms the conclusions of this study and the dependence of ELPs on their system architecture and properties. This also re-opens the question whether the solar system is special or not.\\

The observational bias used in this study is simple and can easily be adapted to any observed system, making it a practical tool. However, years of RV detection have taught us that the observational bias inherent in this method is more complex than a fixed detection threshold. It depends on various factors, such as the star's activity, the presence of other planets in the system, the frequency of observations, and the orbital period. These factors can influence the precision and sensitivity of the detection method, making it a more complex and nuanced process than a simple fixed threshold. A probabilistic observational bias depending on the mass and period of a planet could address this issue, which was studied in \cite{Mayor2011}. This study is also dependent on the model used. Although the diversification of the populations used aims to broaden the predictions, the reliability of these predictions depends on the similarities of the Bern model with the populations of planetary systems in the solar neighbourhood. Testing these predictions on other planetary system populations produced by different global models could either confirm or refute them.\\

To conclude, this work focuses solely on the most easily obtainable parameters of planets (period, radius, and mass). However, future missions such as JWST \citep{Gardner2006}, ARIEL \citep{Tinetti2021}, or later LIFE \citep{LIFE1} will allow access to more complex parameters of planets such as composition, atmospheric structure, water fraction, etc. Using these new parameters in this type of study can prove to be highly valuable and improve the already established correlations between planet properties in a system.

\begin{acknowledgements}
This work has been carried out within the framework of the National Centre of Competence in Research PlanetS supported by the Swiss National Science Foundation under grants 51NF40\_182901 and 51NF40\_205606. The authors acknowledge the financial support of the SNSF. We want to thank the anonymous referee for the valuable comments and suggestions that helped us improve the manuscript, and Angelica Psaridi for her precious help.
\end{acknowledgements}

%
%
\bibliographystyle{aa}
\bibliography{paper.bib}
\begin{appendix}
\onecolumn
\section{Discussion on the exclusion of planets smaller than 0.5 M$_{\oplus}$}\label{App:exclusion}
\begin{figure}[h]
    \centering
    \includegraphics[width=5cm]{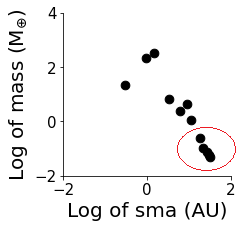}
    \includegraphics[width=5cm]{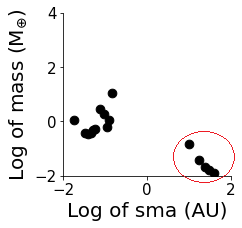}
    \includegraphics[width=5cm]{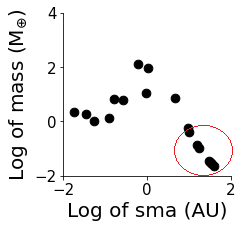}
    \includegraphics[width=5cm]{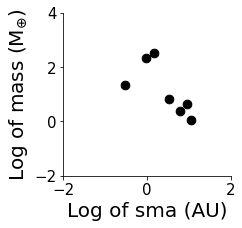}
    \includegraphics[width=5cm]{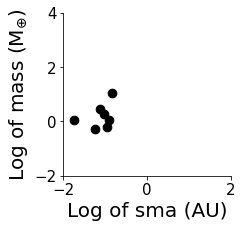}
    \includegraphics[width=5cm]{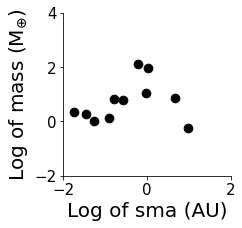}
    \caption{Example of three synthetic systems before (upper panels) and after (lower panels) the selection criterion of M$_p$~>~0.5~M$_{\oplus}$. The red circles highlight the failed planets forming a tail in the colder region of the systems.}
    \label{fig:selection_smallplanets}
\end{figure}

The selection criterion that excludes synthetic planets with a mass lower than 0.5~M$_{\oplus}$ mentioned in Sect. \ref{subsec:bernmodel}  was not chosen arbitrarily and aimed to address a specific issue. In the planetary systems produced by the Bern model, planetary embryos with approximately the size of the moon (0.01~M$_{\oplus}$) are injected before the computation starts. Removing all final planets that are too small (M$_p$~<~0.5~M$_{\oplus}$) essentially means removing planets for which we have not gained much information (a significant part of the final mass is actually from the mass that has been injected initially). These can be referred to as “failed” planets. A large concentration of these failed planets is found in the colder regions of our systems (red circles in the top row of Fig. \ref{fig:selection_smallplanets}), where they form a small, unnatural `tail'. This abundance of small failed planets undoubtedly distorts the architecture of the Bern model systems, and implementing a threshold of 0.5~M$_{\oplus}$ (second row of Fig. \ref{fig:selection_smallplanets}) allows us to get rid of them. Using a smaller threshold would only eliminate some of these small planets. For example, keeping planets of 0.1~M$_{\oplus}$ would mean keeping planets where 10\% of the mass (after 20~Myr) comes from the manually injected planetary embryos at the beginning of the simulation.\\
Although this threshold of 0.5~M$_{\oplus}$ is scientifically coherent, we also studied its impact on results. It can influence this study on the proportion of systems in each architecture and the proportion of systems hosting an ELP in each theoretical and biased architecture.
We compared the effects of a threshold at 0.1 and 0.5~M$_{\oplus}$ on the proportion of systems in each theoretical architecture class and the proportion of systems with an ELP. The \textit{Anti-Ordered}, \textit{Ordered}, and \textit{Mixed} classes are almost unaffected by this change. However, the \textit{Low-mass} and \textit{n~=~1} classes are indeed affected (when the threshold is at 0.1~M$_{\oplus}$, we find 93\% of \textit{Low-mass} systems and 6.3\% of \textit{n~=~1} systems, for example, compared to 75\% and 24\% when the threshold is at 0.5~M$_{\oplus}$).
Even though these proportions change (less than 20\% of the systems nonetheless), the proportions of biased architectures do not change. The proportions of systems in each architecture class, and the proportions of systems with ELPs in each architecture class remain the same within 0.1\% between a threshold of 0.1 and 0.5~M$_{\oplus}$. Indeed, the observational bias (very strict for lateM-pop, 1.55~m/s) already removes all planets whose mass is less than 0.5~M$_{\oplus}$. This shows that the threshold of 0.5~M$_{\oplus}$ is coherent for this study.

\section{Discussion on the selection of max(M$_p$), S(C$_1$), and V(C$_2$) values and their robustness}\label{App:bornes}
In Sect. \ref{subsubsec:framework}, we present the limit values chosen for the quantities M$_p$, S(C$_1$), and V(C$_2$). We will discuss this selection and the robustness of the values chosen for these two coefficients in more detail in this appendix. The value chosen for these quantities could theoretically affect the distribution of synthetic systems in the architecture classes and the proportion of systems in each with and without ELP.

\subsection{max(M$_p$)}
\begin{figure}[h]
    \centering
    \includegraphics[width=12cm]{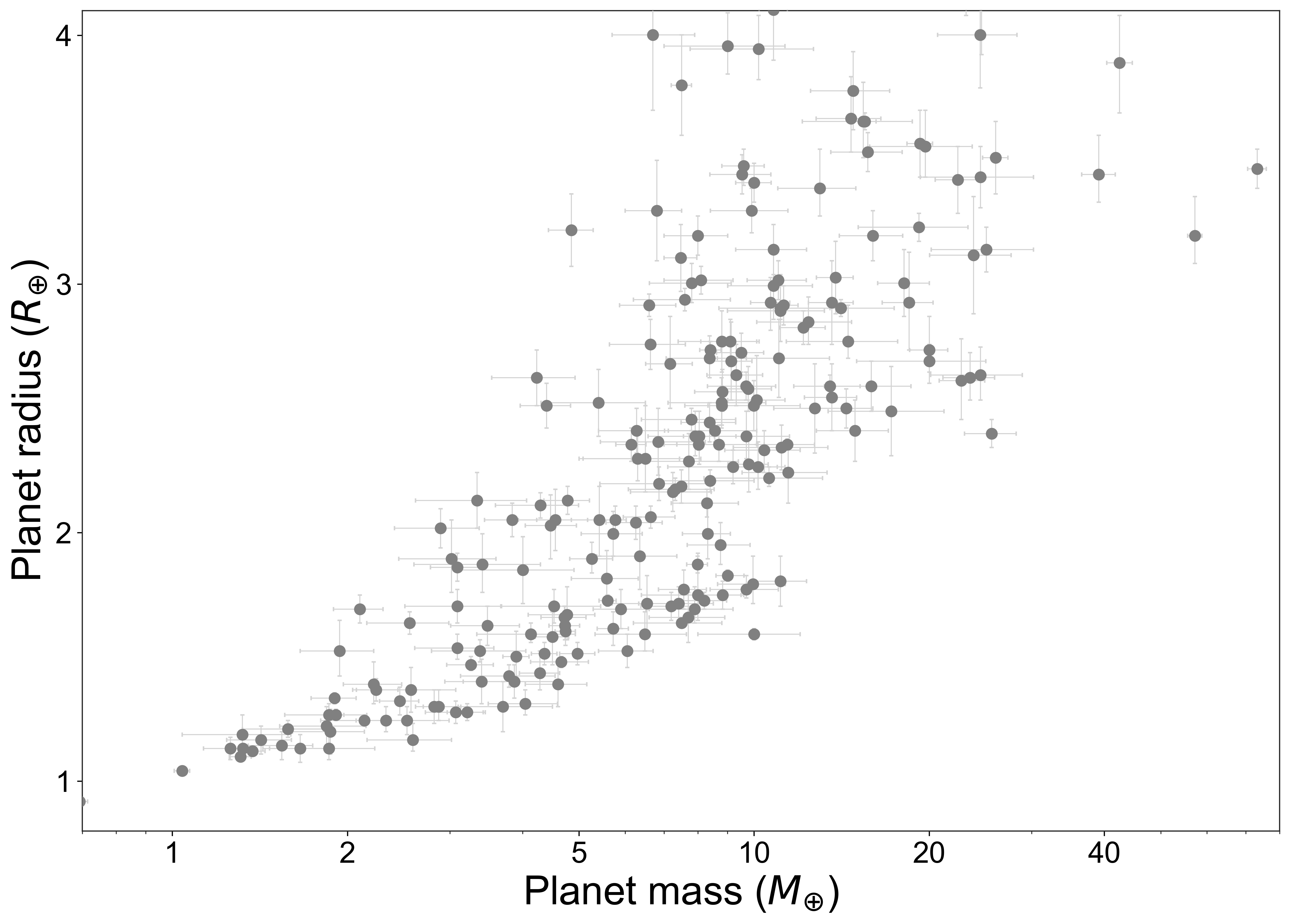}
    \caption[]{Mass-radius diagram of small exoplanets (radii up to 4~R$_{\oplus}$) with well-constrained densities orbiting FGK stars from the PlanetS catalog\protect\footnotemark \citep{otegi2020,parc2024}. Figure adapted from \cite{Psaridi2024}.}
    \label{fig:mass-radius-psaridi}
\end{figure}

The parameter max(M$_p$) represents the mass limit that a system's most massive planet can take, separating the \textit{Low-mass} class from the other three architecture classes. The 20~M$_{\oplus}$ value was not chosen randomly. Indeed, \cite{Bodenheimer1986} define the critical mass M$_{crit}$ of a planetary body, which separates accretion in the regular regime from a runaway regime between 10 and 30~M$_{\oplus}$. We see minimal variation in the distribution of the synthetic populations in the three architecture classes, and in the proportion of systems with or without ELPs in each class when changing the value of 20~M$_{\oplus}$ by 10 or 20\%. However, variation is observed when extending it to the entire range of values of the `critical mass' defined by \cite{Bodenheimer1986}, from 10 to 30~M$_{\oplus}$, but this central value also corresponds to an observed upper limit of the `small planet' regime. Indeed, in the Fig. \ref{fig:mass-radius-psaridi} that represents a sample of small observed planets in a mass-radius diagram, we see that the boundary of the `small planet' regime (radius < 4~R$_{\oplus}$) is approximately between 20~M$_{\oplus}$ and 30~M$_{\oplus}$. 10~M$_{\oplus}$ is in the middle of this regime, and 30~M$_{\oplus}$ is already slightly outside. However, even changing this limit from 20 to 30~M$_{\oplus}$ gives us a variation of a maximum of 3.8\% for theoretical architecture and 5\% for biased architecture in the distribution of synthetic systems in the five architecture classes and proportion of systems with or without ELPs. Us, 20~M$_{\oplus}$ is the most sensible value.
\footnotetext{Catalog available on the Data \& Analysis Center for Exoplanets (DACE) platform (\url{https://dace.unige.ch})}

\subsection{S(C$_1$) and V(C$_2$)}
\begin{figure}[h]
    \centering
    \includegraphics[width=6cm]{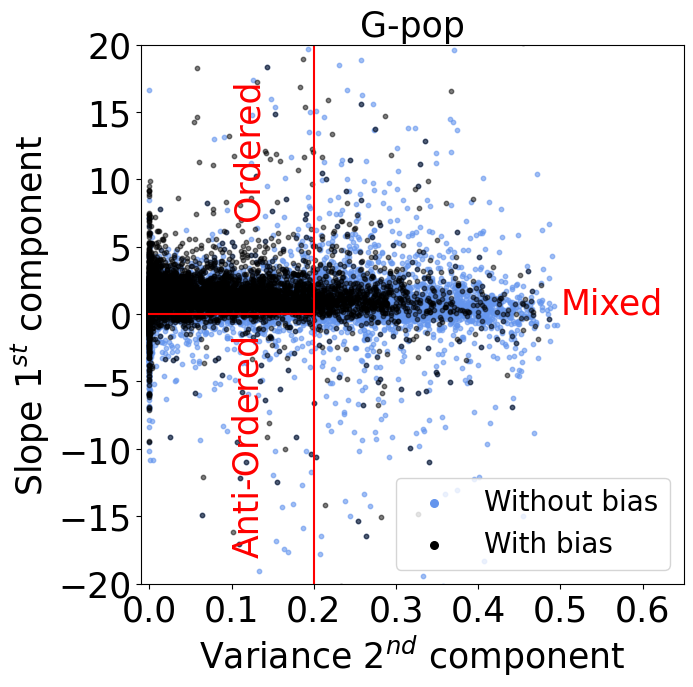}
    \includegraphics[width=6cm]{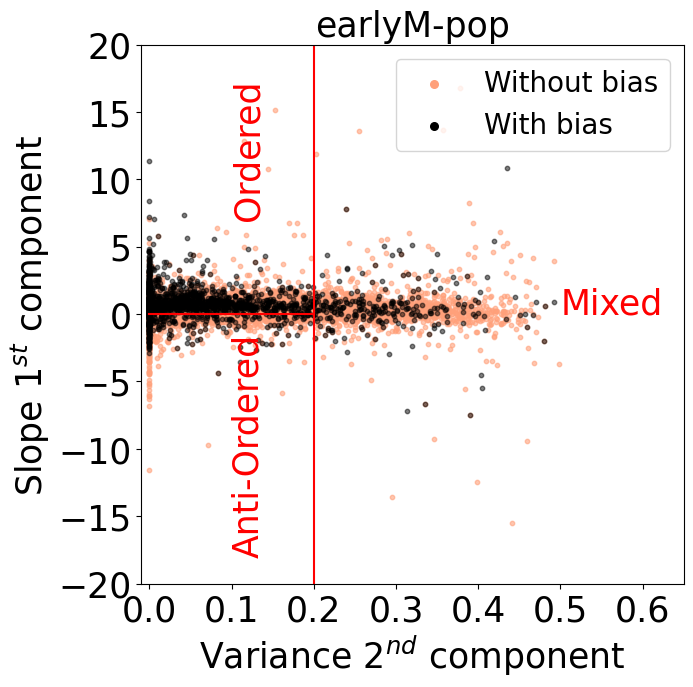}
    \includegraphics[width=6cm]{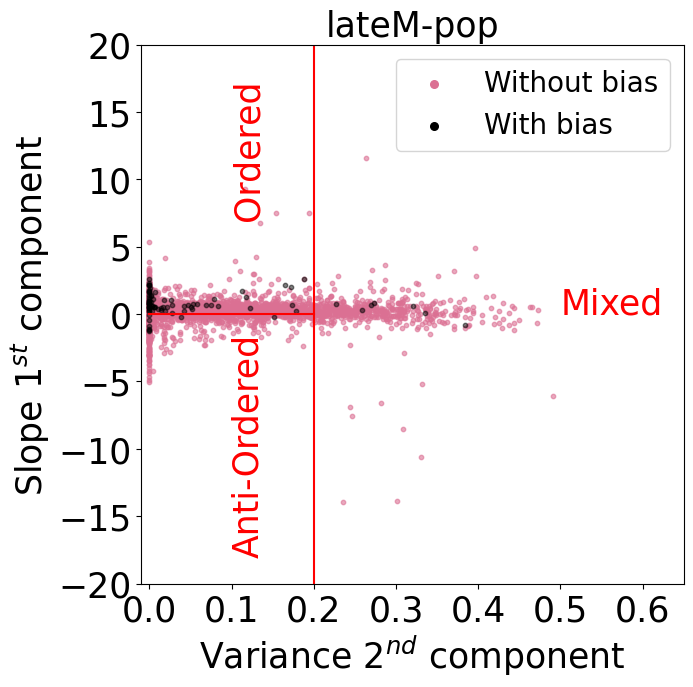}
    \caption{Three populations of synthetic planetary systems with at least two planets and one planet larger than 20 M$_{\oplus}$ displayed in V(C$_2$)-S(C$_1$) diagrams.}
    \label{fig:cs_cv}
\end{figure}

\begin{figure}[h]
    \centering
    \includegraphics[width=6cm]{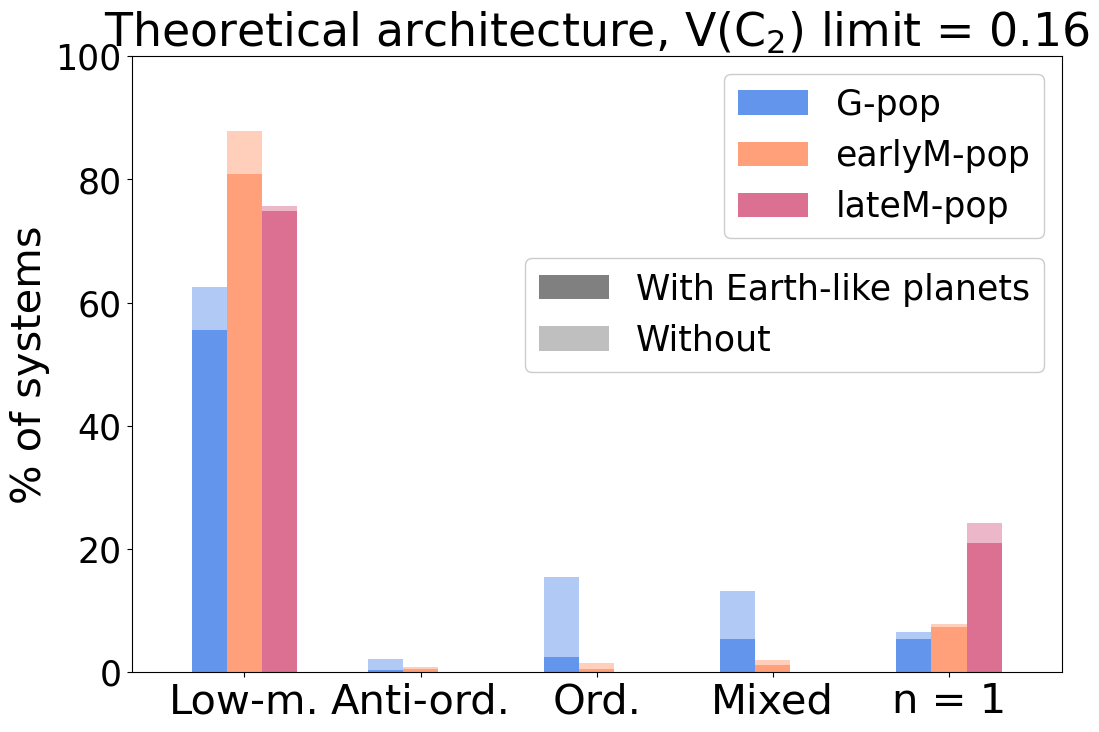}
    \includegraphics[width=6cm]{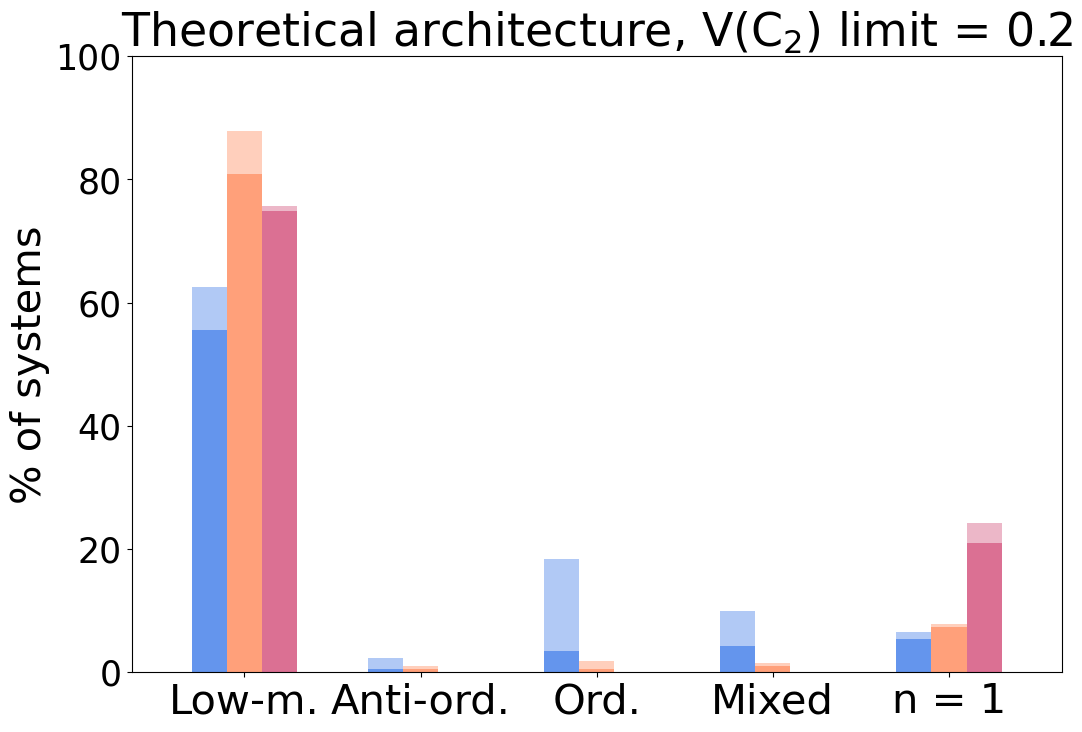}
    \includegraphics[width=6cm]{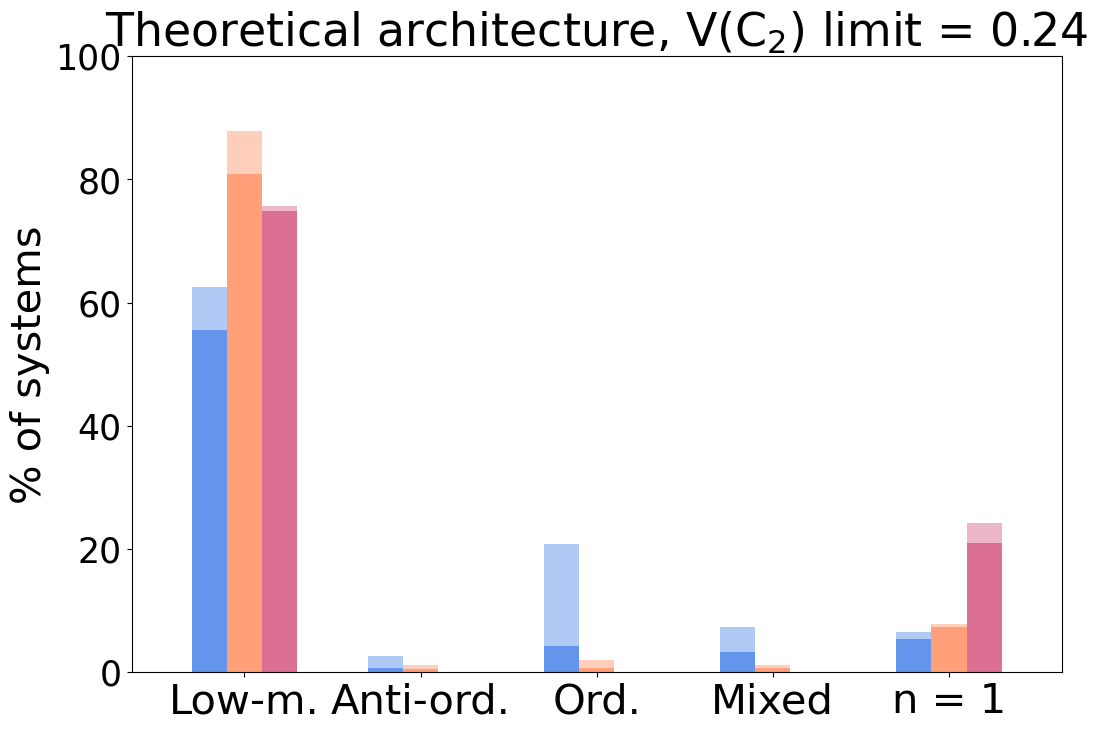}
    \includegraphics[width=6cm]{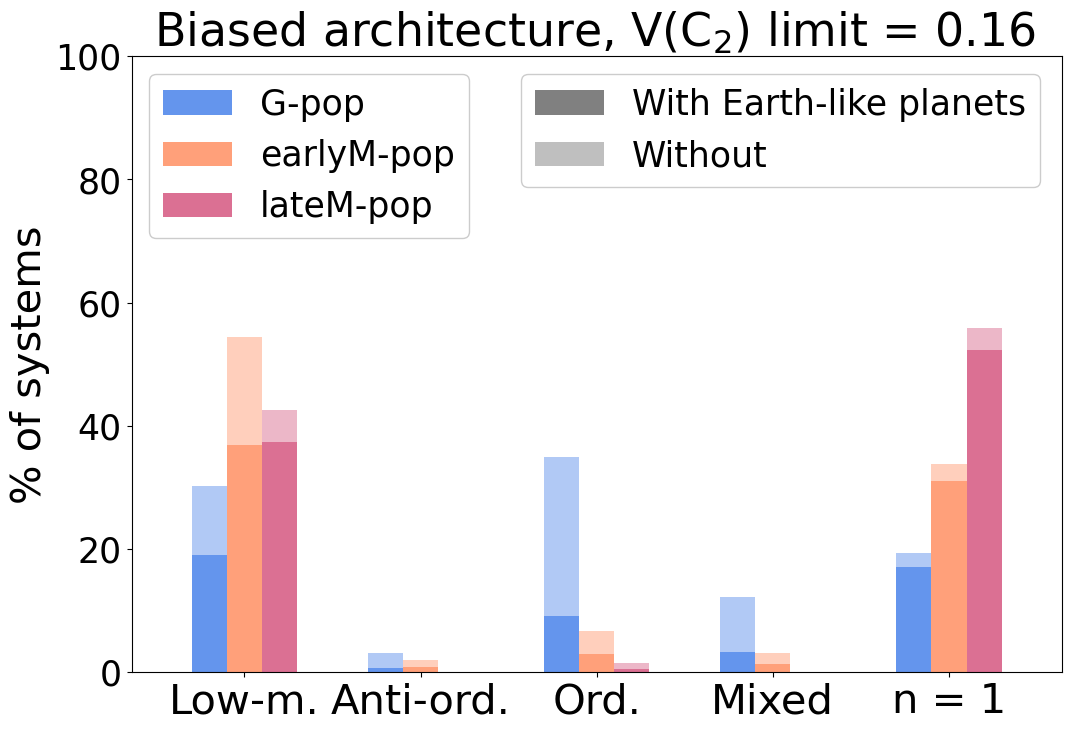}
    \includegraphics[width=6cm]{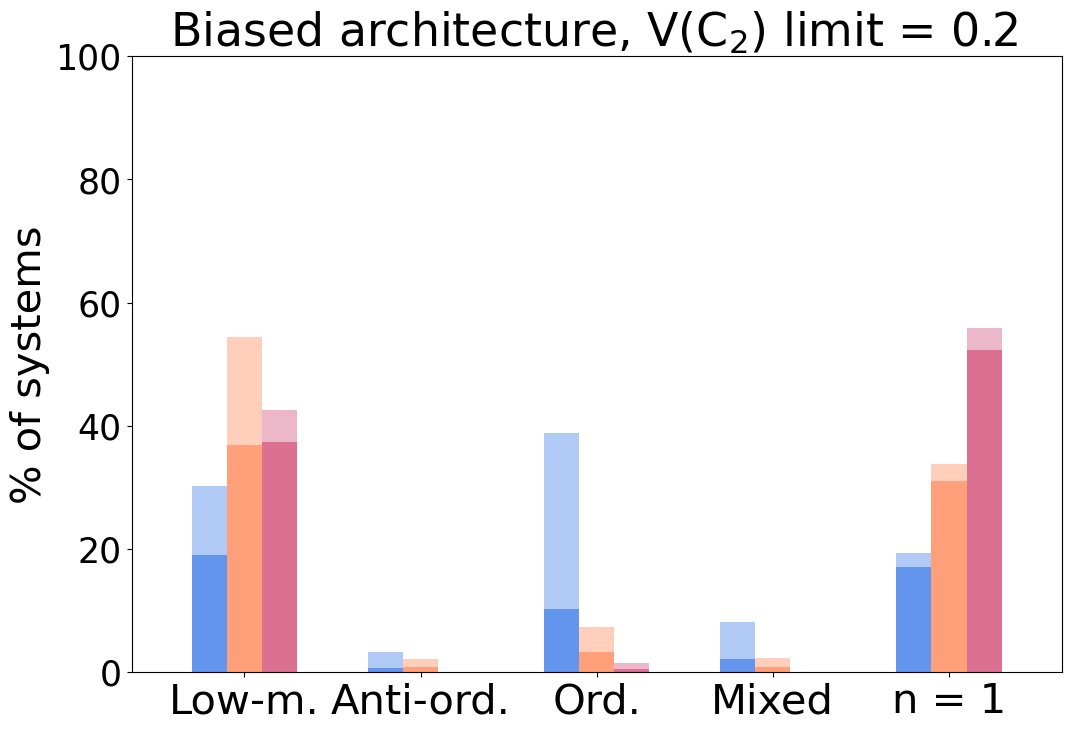}
    \includegraphics[width=6cm]{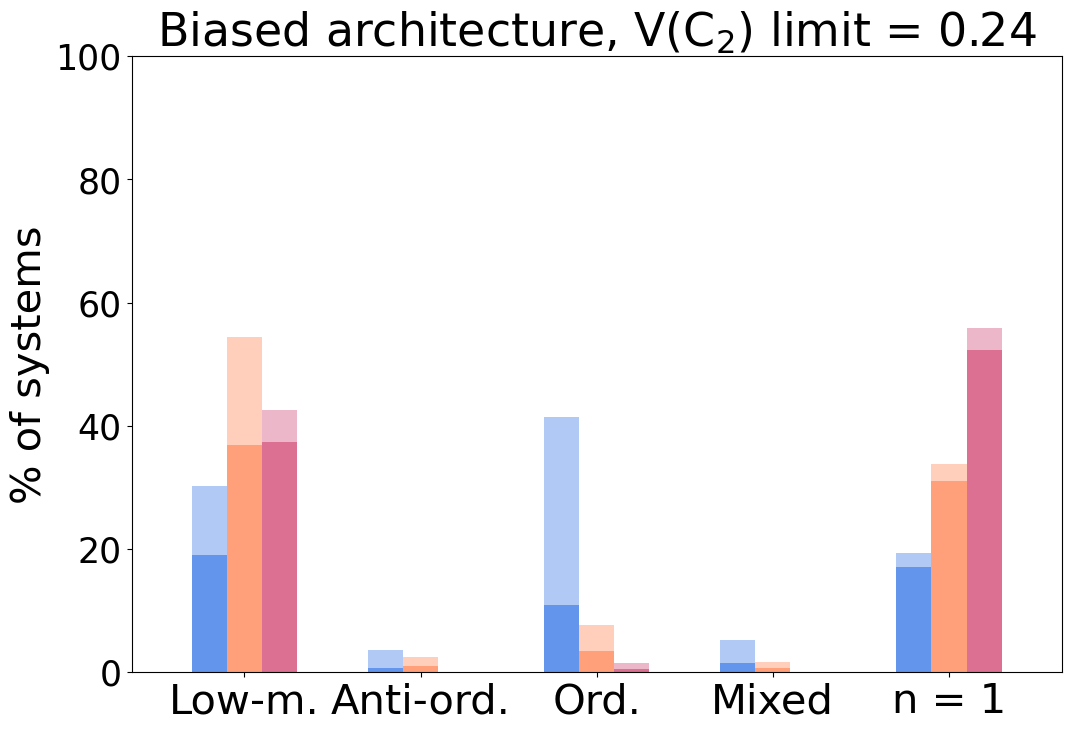}
    \caption{Distribution in different architecture classes of the synthetic planetary systems of the three populations by varying the limits of V(C$_2$). Upper panels correspond to theoretical architecture and lower panels correspond to biased architectures. }
    \label{fig:variationVC2}
\end{figure}

Figure \ref{fig:cs_cv} represents the three populations of synthetic theoretical and biased planetary systems with at least two planets and one planet larger than 20~M$_{\oplus}$, in a V(C$_2$)-S(C$_1$) diagram. The coloured points correspond to complete systems (theoretical architectures). In contrast, the black points correspond to biased systems (biased architectures). The slope limit of the first component separates the `increasing' and `decreasing' systems, which we call \textit{Ordered} or \textit{Anti-Ordered}, and is a definition that can be considered `immutable' (it does not make sense to classify a system as \textit{ordered} if the slope of the main component is negative, for example). \\
Figure \ref{fig:variationVC2} shows the distribution of the three populations of synthetic systems in the five architecture classes when the value of V(C$_2$) is varied by $\pm$ 20\% around its nominal value of 0.2. The panels in the first row correspond to the theoretical architectures and the panels in the second row correspond to the biased architectures. The limit value of V(C$_2$) increases from left to right.\\
The limit on the variance of the second component V(C$_2$), initially calibrated by eye to match what the authors deemed to be a system with high variability or not, corresponds to about half the range of values taken by V(C$_2$) ($\sim$ 40\%) (see Fig. \ref{fig:cs_cv}). In addition, if we observe the distribution of populations in the architecture classes (Fig. \ref{fig:variationVC2}, we can see that changing the value of V(C$_2$) by 20\% does not dramatically change the proportions of classes. Indeed, there are more \textit{Ordered} systems and fewer \textit{Mixed} systems as we increase V(C$_2$) among theoretical architectures (top row) and biased architectures (bottom row), but this difference is barely a few percents. In theoretical architectures, the largest difference in the proportion of systems in an architecture is 3.2\% and for biased architectures, it is 4.1\%. So, although there is no distinct natural value in Fig. \ref{fig:cs_cv}, we can conclude that the architectural classes are robust.

\end{appendix}

\end{document}